\documentstyle[12pt]{article}
\newcommand{\bq}{\begin {eqnarray} & \displaystyle}
\newcommand{\eq}{ & \end {eqnarray}}
\newcommand{\ra}{\rightarrow}
\newcommand{\la}{\leftarrow}
\newcommand{\se}{\simeq}

\newcommand{\bi}{\bibitem}

\begin{document}

\thispagestyle{empty}

\setcounter{page}{0}

\vspace*{2cm}
\begin{center}
{\Large\bf Calculation of the D and B meson lifetimes} \end{center}

\begin{center}{\bf Victor Chernyak}
\footnote{e-mail: {\bf CHERNYAK@INP.NSK.SU}}
\end{center}

\begin{center}{\bf Budker Institute of Nuclear Physics,\, 630090
 Novosibirsk-90,\, Russia} \\  and \\
 {\bf Novosibirsk State University} \end{center}

\vspace*{2 cm}
\begin{center} {\bf Abstract} \end{center}

Using the expansions of the heavy meson decay widths in the heavy quark mass
and QCD sum rules for estimates of corresponding matrix elements,\, we
calculate the $D^{\pm,o,s}$ decay widths and the $B^{\pm,o,s}$ lifetime
differences. The results for D mesons are in agreement with the data,\,
while it is predicted that $[\Gamma (B^o)-\Gamma (B^-)]/\Gamma_B\se 4\%\,,$
and the lifetime difference of the $B^o$ and $B_s$ mesons is even smaller.
The role of the weak annihilation and Pauli interference contributions to the
lifetime differences are described in detail. In the course of self-consistent
calculations the values of many parameters crucial for calculations with
charmed and beauty mesons are found. In particular,\, the perturbative pole
quark masses are: $M_c\se 1.65\,GeV,\,\, M_b\se 5.04\,GeV\,,$ and the
decay constants are: $f_D(M_c)\se 165\,MeV\,,\,\,f_B(M_b)\se 113\,MeV\,$.
It is also shown that the nonfactorizable corrections to the $B^o-{\bar B}^o$
mixing are large,\, $B_B\se (1-18\%)\,.$ The values of the unitarity
triangle parameters are found which are consistent with these
results and the data available (except for the NA31 result for the $\epsilon
^{\prime}/\epsilon$ which is too large): $\lambda\se 0.22\,,\,\,
A\se 0.825\,,\,\,\rho \se -0.4\,,\,\, \eta \se 0.2\,.$

\newpage
\renewcommand{\thefootnote}{\arabic{footnote}}
\setcounter{footnote}{0}


 {\bf 1\,. \quad  Introduction}\\

It is a long standing challenge for theory to calculate the D and B meson
decay widths. On the qualitative side,\, two mechanisms were invoked to
explain the pattern of the D meson lifetime differences: weak annihilation
(WA) \cite {FM},\,\cite{BWS},\,\cite{BSS},\, and Pauli interference (PI)
\cite{Gub},\,\cite{VS0},\,\cite{BGT},\,\cite{VS}. As for WA,\, it was expected
that because an admixture of the wave function component with an additional
gluon or the emission of a perturbative gluon,\, both
remove a suppression due to helicity conservation (which leads to $Br\,
(\pi\ra e\,\nu)/Br\,(\pi\ra \mu\,\nu)\sim 10^{-4})$,\, the $D^o$ meson decay
width is enhanced. On the other hand,\, it was expected that the
destructive Pauli interference of two d-quarks (the spectator
and those from a final state) suppresses the $D^{\pm}$ meson decay width.
As for WA,\, there were no reliable
calculations at all. For PI,\, simple minded estimates (see sect.8)
give too large an effect which results in a negative $D^{\pm}$ decay width.

For B mesons,\, it was clear qualitatively that all the above effects
which are of a pre-asymptotic nature and die off at $M_Q\ra \infty$,\, will be
less important. However,\, because the patern of the D meson lifetime
differences was not really explained and well understood,\, this prevented to
obtain reliable estimates of the B meson lifetime differences,\, and only
order of magnitude estimates are really available: $[\delta \Gamma(B)/\Gamma_
B]:[\delta \Gamma(D)/\Gamma_D]\sim O(f_{B}^{2}M_{c}^{2}/f_{D}^2M_{B}^{2})
\sim O(10^{-1})$.

Moreover,\, as for WA contributions through perturbative gluon
emission (which is formally a leading correction $\sim
O(\Lambda_{QCD}/M_Q)$ to the deacay width
and was expected before to be potentially the most important),\, it has
been emphasized recently \cite{BU}
that such contributions are of no help at all
because,\, being large (at least formally at $M_Q \ra \infty$) term by term,\,
they cancel completely in the inclusive widths,\, both $O(1/M_Q)$ and $O(
1/M_Q^2)$ terms. \footnote{ The separate perturbative WA contributions into the
total width have the form: $\delta \Gamma/\Gamma_{Born}\sim f_D^2/
\epsilon_o^2\,,\,\,\, f_D^2\sim \mu_o^3
/M_Q,\,$ where $\epsilon_o$ is the binding
energy of the spectator quark in the meson. So,\, they are highly infrared
sensitive and singular at $\epsilon_o\ra 0.$
It seems clear that there can not be power infrared singularities in
the inclusive decay width,\, so that it is not so
surprising that all such terms cansel (i.e. the terms $\sim
O(1/\epsilon_o^2)$ and $\sim O(1/\epsilon_o)$,\, but
not $\log \epsilon_o$ which describes hybrid log.\,) .}
It will be shown below (see sect.9) that,\, nevertheless,\, there
are important WA contributions but on the nonperturbative level.

Considerable progress has been achieved recently in applications of the
operator product expansion to the calculation of the heavy meson decay width.
In particular,\, it was shown that there are no $O(1/M_Q)$ corrections to
the Born term and first nonperturbative corrections $O(1/M_Q^2)$
were calculated explicitly \cite{BUV1},\,\cite{BS1}. However,\,
these contributions are all nonvalence (i.e. one and the same for all
$D^{\pm,o,s}$ mesons),\, and so have nothing to do with lifetime
differences. They are important however for the calculation of the absolute
decay rates. Recently a number of papers appeared \cite{LS},\,\cite{Big},\,
\cite{LN},\,\cite{SUV},\, where these results were applied to determine the
values of the quark masses,\, $M_c$ and $M_b$,\, and $|V_{cb}|.$

The purpose of the present work was to calculate
the D and B meson decay widths,\, with the main emphasis
on the calculation of lifetime differences. It is shown that the
 $D^{\pm,o,s}$ lifetimes can be calculated with a reasonable accuracy,\,
and concrete predictions for the $B^{\pm,o,s}$ meson
lifetime differences are given also.

The scope of the paper is as follows:
1)\,\, Introduction\\
2)\,\, Definition of the heavy quark mass\\
3a)\,  General formulae:\, c-quark\\
3b)\,  General formulae:\, b-quark\\
4)\,\, $D\ra e\,\nu+X$. Determination of $M_c$\\
5)\,\, Mass formulae. Determination of $M_b$ and $\bar \Lambda$\\
6)\,\, $B\ra e\,\nu+X$. Determination of $|V_{cb}|$\\
7)\,\, Calculation of $f_D,\, f_B$\\
8)\,\, Difficulties with naive estimates\\
9)\,\, Nonfactorizable contributions: gluon condensates\\
10)\, Nonfactorizable contributions: quark condensates\\
11)\, Corrections to semileptonic widths\\
12)\, $\lambda$ and $\bar \lambda$\\
13a)\, Calculation of $D^{\pm,o,s}$ decay widths\\
13b)\, Calculation of $B^{\pm,o,s}$ lifetime differences\\
14)\, $B^o-{\bar B}^o$ mixing\\
15)\, The unitarity triangle\\
16)\, Summary and conclusions\\

{\bf 2.\quad  Definition of the heavy quark mass}\\

In what follows,\, expansions in powers of the heavy quark mass
and some formulae of the HQET (Heavy Quark Effective Theory)
(see,\, i.e. the reviews \cite{IW},\,\cite{Ge},\,\cite{Neu}\,)
are used. In the standard approach,\, the perturbative pole mass, $M_p$,\,
is chosen as the HQET expansion parameter.
The perturbative pole mass is,\, clearly,\, a distinguished parameter
because it is scheme and gauge independent. So, it is natural that all
calculations of observable quantities are expressed usually through $M_p$.

It was pointed out recently \cite{BSUV},\,\cite{BB},\,\cite{BBZ} that the
perturbative pole mass is an ill-defined quantity and contains an intrinsic
uncertainty $\sim O(\Lambda_{QCD})$,\, because the perturbation theory series
for it diverges due to renormalon effects.

Analogous renormalon divergences are well
known and originate from the fact that there is an internal scale in the
theory: $\Lambda_{QCD}$- the position of the coupling constant infrared
pole. So,\, this scale reveals itself when perturbative loops
are integrated up to $k_i\ra 0$ (as in the case with the pole mass). But
because $\Lambda_{QCD}$ can not appear
explicitly at any order of renormalized perturbation theory,\, the
perturbative series diverges. These infrared singularities are
cured usually by nonperturbative contributions,\, and it seems most people
are not worrying especially about them.

To avoid from the beginning the infrared region contributions,\, it was
proposed in \cite{BSUV} to use some infrared safe mass parameter,\,
$M_{\mu}$,\, instead of $M_p$. As for a concrete choice of $M_{\mu}$,\,
two variants are considered usually. The first one (in a
spirit of the Wilson operator expansion) is to calculate all loops by cutting
out the $k_i\leq \mu$ region contributions. A deficiency of this
variant is that it is practically impossible to perform such calculations.
The second one is to use the current quark mass normalized far off mass shell
which suppresses the infrared region contributions. Because this current mass
differs from $M_p$ already at the first loop level,\, this variant spoils
finally all the usefullness of the standard HQET approach,\, replacing the
standard decomposition in powers of $1/M_p$ by the usual perturbation theory
decomposition in powers of $1/\log M_{\mu}$. Besides,\, because the current
quark mass is renormalization scheme and gauge dependent,\, all this is highly
inconvenient.

In connection with this, we describe below another redefinition of $M_p$
and introduce the "hard pole mass",\, $M_o$,\, which is free of renormalon
singularities by definition and differs from $M_p$ by only the term
$\sim \Lambda_{QCD}$ (not by $\sim \alpha_s M_p\,)$.
We emphasize that,\, as far as we do not calculate explicitly
high order perturbation theory corrections to observable quantities,\, the
whole problem of the renormalon is of abstract interest only and can be
formally solved on a "hand waving level",\, i.e. by some redefinitions only,
\, without making any real changes in all the formulae available.

The above  "hard pole mass",\, $M_o$,\, can be
connected with a definition of the matrix element of the local operator.
Although it is clear beforehand,\, let us demonstrate explicitly,\, using
the example considered in \cite{BSUV},\, that the
relevant operator to use for a redefinition of $M_p$ is $H_{light}$,\, i.e.
the light degrees of freedom part of the Hamiltonian.
With this purpose,\, let us consider the Hamiltonian and take
its average over the heavy meson state,\, $|M_H>:$
\bq <M_H|\,H_{tot}\,|M_H>=M_\mu <M_H|\,({\bar Q}\,Q)_\mu\,|M_H>+ \nonumber \eq
\bq <M_H|\,\frac{1}{2}
({\vec E}^2+{\vec H}^2)_\mu|M_H>+<H_q>+O(\mu^2/M)\,,\eq
where $\mu$ serves as an upper cut off in the matrix elements of operators
and as lower cut off in $M_\mu$,\, $H_q$ is the light quark Hamiltonian,\,
$<M_H|({\bar Q}Q)_\mu|M_H>=1+O(\mu^2/M^2)\,.$

In the example considered in \cite{BSUV} (in the static limit):
\bq M_\mu=M_p-\delta M, \quad \delta M=-i {{16\pi}\over{3}}\int {{{d^4}k}
\over{(2\pi)^4}}\,\alpha_s\,\phi(k)\,, \nonumber \eq
\bq \phi(k)={{1}\over{k_o+i\epsilon}}\,{{1}\over{k^2+i\epsilon}}\,{{\mu^2}
\over{\mu^2-k^2-i\epsilon}}\,,\eq
and when $\alpha_s$ in Eq.(2) is substituted by running $\alpha_s(k^2)$
in the form:
\bq \alpha_{s}({\vec k}^2)=\alpha_{s}(\mu^2)\sum_{n=0}^{\infty}\left (\frac{b_o
\,\alpha_{s}(\mu^2)}{4\pi}\log \frac{\mu^2}{{\vec k}^2}\right )^n\,,  \eq
this leads to a divergent perturbative series:
\bq \delta M=\mu\,\sum_{n}^{\infty}{\alpha_s}^n\,C_n\,,\quad C_n
\sim (\frac{b_o}{2\pi})^n\,n!\,\,.\eq
 On the other hand,\, it is
not difficult to see that the insertion of the vertex ${\vec E}^2/2$ into
the same diagram which gives $\delta M,\,$ adds the contribution:
\bq <M_H|\,{{1}\over{2}}{\vec E}^2\,|M_H>=-i{{16\pi}\over{3}}\int{{{d^4}k}
\over{(2\pi)^4}}\,\alpha_{s}\,\phi(k)
\left (\frac{-{\vec k}^2}{k^2+i\epsilon}\, \right )= \delta M\,,\eq
which cancels the renormalon contribution in Eq.(1 ) and leaves the pole
mass $M_p$ instead of $M_\mu$. \footnote {This is natural
because the whole answer has to be independent of $\mu$,\, and considering
formally the case when there are no nonperturbative contributions and the
matrix element is taken over the on mass shell heavy quark state, the total
answer is $M_p$\,.}

Let us proceed now with a standard on mass shell renormalization scheme and
rewrite Eq.(1 ) in the form:
\bq <M_H|\,H_{tot}\,|M_H>=M_p <M_H|\,{\bar Q}\,Q\,|M_H>+<M_H|
\,H_{light}|M_H>+ & \nonumber \\ & \displaystyle
\frac{2\,\delta {\tilde m}^2}{M}+O(\Lambda_{QCD}^3/M^2)\,,\eq
where \cite{BUV1} (see sect.5 for the explicit form of $\delta
{\tilde m}^2$):
\bq <M_H|\,{\bar Q}Q\,|M_H>=1-{\delta {\tilde m}^2}/M^2+
O(\Lambda_{QCD}^3/M^3)\,,\quad \delta {\tilde m}^2=O(\Lambda^2_{QCD})\,.\eq
It is not difficult to check that:
\bq <M_H|\, H_{light}\,|M_H>= \Lambda_o \left (1+O(\Lambda_
{QCD}^2/M^2)\right )\,,\eq
where $\Lambda_o$ is a finite number $\sim \Lambda_{QCD}$,\, independent of M.
Clearly,\, it can be understood as the light quark self energy plus the
binding energy.

Now,\, let us redefine the quark pole mass,\, $M_p$,\, and introduce the
"hard pole mass",\, $M_o$\,:
\bq M_p=M_o\left (1+\Delta_1+\Delta_2+\cdots \right )\,,\eq
where $\Delta_1$ is defined formally,\, for instance,\, as the leading
asymptotic part of the divergent perturbative series for $M_p$:
\bq \Delta_1=\sum_{n=N_1}^{\infty}\alpha_{s}^{n}(M_o)\,C_n^{(1)}=A_1\frac
{\Lambda_{QCD}}{M_o}\,,\quad C_n^{(1)}\sim \left (\frac{b_o}{2\pi}
\right )^n\,n!\,,\eq
and $A_1\sim (1+O(\alpha_s)),\, N_1\se 1/\alpha_s(M_o)$. \footnote{ More
precisely,\, if any meaning can be given to the divergent sum in Eq.(10) ,\,
it can also give higher order terms $\sim\Lambda_{QCD}^2/M_o^2+...$,\,
but these can be included into a redifinition of $\Delta_2,\,$ etc.}
$\Delta_2$ in Eq.(9)
is defined analogously through a subleading part of the series,\, so that
$\Delta_2=A_2\,\Lambda^2_{QCD}/M_o^2$,\, etc. So,\, by definition,\, $M_o$
in Eq.(9) is free of divergent parts of renormalon contributions. What
is important,\, is that the "bad" quantity $M_p$ and the "good" quantity
$M_o$ differ not by $\sim \alpha_s(M_o)M_o$ terms,\, but terms
$\sim \Lambda_{QCD}$ only.

Clearly,\, we can absorb the term $\Delta_1$ into a redefinition of
$\Lambda_o$,\, the term $\Delta_2$ into a redefinition of $\delta
{\tilde m}^2$,\, etc.,\, so that Eq.(6) can be rewritten as:
\bq <M_H|\,H_{tot}\,|M_H>=M_o <M_H|\,{\bar Q}\,Q\,|M_H>+{\bar \Lambda}
+\frac{2\,\delta m^2}{M_o}+O(\Lambda_{QCD}^3/M_o^2)\,,\eq
\bq {\bar \Lambda}=\Lambda_o+\Delta_1 M_o\,,\quad \delta m^2=
\delta {\tilde m}^2+\Delta_2 M_o^2\,.\eq

In connection with the above,\, let us point out the following.
Strictly speaking,\, because the real expansion parameter is
$\Lambda_{QCD}/M$,\, taking account of higher order terms of the
perturbative series in $\alpha_{s}(M)$
is not justified without simultaneously taking account of power corrections
in $\Lambda_{QCD}/M$.
So,\, even if we have a convergent series: $\sum \alpha_{s}^{n}(M) B_n$,\,
the term $\alpha_{s}^{N_1}(M)\,B_{N_1}$ becomes $O(\Lambda_{QCD}/M)$
starting with $N_1$ (depending on the behaviour of $B_n$ at large n);\,
starting with $N_2$ the term $\alpha_{s}^{N_2}(M)\,B_{N_2}$ becomes
$O(\Lambda_{QCD}^2/M^2)$,\, etc. For instance,\, for $B_n\sim 1$ at large n:
$N_1\sim [\alpha_s(M)\,\log 1/\alpha_{s}(M)]^{-1}$,\, while for $B_n\sim n!:
N_1\sim [\alpha_s(M)]^{-1}$, which is only slightly larger than those for
the convergent series. Therefore,\, in any case,\, the account of distant
terms of the perturbative series has to be made only simultaneously with
account of power corrections,\, and we can always transfer these distant
perturbative terms into a redefinition of power corrections (remembering
especially that we are unable to calculate directly these nonperturbative
corrections at present).

Let us turn now to the heavy meson decay width which can be represented in
the form \cite{BUV1}:
\bq \Gamma \sim M_p^5 \left [\,F \Bigl (\alpha_s(M_p) \Bigr )+
O({\Lambda_{QCD}^2}/M^2_p)\right ]\,,\eq
where $F(\alpha_s)$ represents a perturbative series,\, and power
corrections in Eq.(13) are explicitly expressed through
the matrix elements of higher dimension operators (see sect.3).

As was emphasized in \cite{BSUV},\, a new element in the example considered
is that the renormalon effects lead to a contribution $\sim \Lambda_{QCD}$
in the pole mass,\, while there are no nonperturbative corrections to
$\Gamma$ at this level. Correspondingly, the perturbative series for
$F(\alpha_s)$ is also divergent,\, and only a product of them in Eq.(13)
contains no $\sim \Lambda_{QCD}/M$ renormalon contributions.

Now,\, let us reexpress the pole mass in Eq.(13) through the formally defined
"hard pole mass",\, $M_o$\,. Then in the product: $(1+5\,\Delta_1+O(
\Delta_2))\,F(\alpha_s(M_o))$ (where $\Delta_1$ is understood as the divergent
series Eq.(10)),\, the leading divergences cancel,\, so that there is no
$\Lambda_{QCD}/M_o$ correction. As for subleading divergencies which give
corrections $\sim \Lambda^{2}_{QCD}/M_o^2$, they will be cured in the usual way
by explicit nonperturbative matrix elements in Eq.(.13)\,, etc.

The final result of all the above manipulations can be formulated as follows.\\
{\bf 1}. We can use the standard perturbative pole mass,\, $M_p$,\, in the
HQET to obtain usual expansions in powers of $1/M_p$,\, putting no attention
at the time that it is not a "good" quantity.\\
{\bf 2}. When dealing with formulae for directly observable quantities,\, we
can reexpress $M_p$ through the "hard pole mass",\, $M_o$,\, which is a
"good" quantity. All renormalon divergencies will be either explicitly
canceled in perturbative expansions,\, or absorbed by redefinitions of the
original
nonperturbative matrix elements. The net effect will be that all the
original formulae of the $1/M$ expansions will formally stay intact after
all the above redefinitions. Moreover,\, if we calculate explicitly only
a few lowest order loop corrections,\, which is the case usually,\, we
need not even explicit expressions for the divergent tails of perturbative
expansions. It is sufficient to have in mind that all the divergent
contributions are cured by redefinitions.\\
{\bf 3}. What is really important,\, is that (having in mind all the above
redefinitions) we can use now our formulae with only a few first loop
corrections explicitly calculated for a comparison with the data,\, not
worrying much that there may be numerically large corrections from distant
terms of the perturbative series. \\

{\bf 3a. \quad  General formulae:\, c-quark}\\

  The weak Hamiltonian used in what follows has the form ($\,\Gamma_\mu=
   \gamma_\mu (1+\gamma_5)\,)$:
   \bq H_{W}^{c}(\mu=M_c)={{G_F}\over{\sqrt 2}}\,V_{c s}\,V^{*}_{u d}\,
   \left \{\,C_{1}(M_c)\,{\bar s}\,\Gamma_\mu c \cdot {\bar u}\,\Gamma_\mu d+
   C_{2}(M_c)\, {\bar u}\,\Gamma_\mu c \cdot {\bar s}\,\Gamma_\mu d\, \right \}
   + \cdots, \eq
   where the dots denote corresponding Cabibbo-suppressed contributions.
   The coefficients $C_{\pm}=(C_1\pm C_2)/2$ are determined by:
   \bq C_{\pm}(\mu^{2})=C_{\pm}^{L}(\mu^{2})\,\left\{1+{{\alpha_s(\mu^{2})-
   \alpha_s(M^2_W)}\over{\pi}}\,\rho_{\pm}\right \}\,, \eq
   \bq C_{-}^L=\left ({{\alpha_s(\mu^{2})}\over{\alpha_s(M^2_W)}}\right )^
   {4/b_o}\,,\quad C_{+}^L(\mu^{2})=1/{\sqrt {C^{L}_{-}(\mu^{2})}}\,,\eq
   and the values of $\rho_{\pm}$ can be found in \cite{A1}.
   We use below in our calculations:
   \bq {{d\,\alpha_{s}(t)}\over{d\,t}}=\beta(\alpha_{s})=-{{{b_{o}}
   \over{4\,\pi}}}
   \,\alpha_{s}^2\,(1+\Delta\,\alpha_{s})\,, \eq
   \bq \Delta={{1}\over{4\,\pi}}\,{{b_1}\over{b_o}}\,,\quad
   b_{o}=(11-{{2}\over{3}}\,n_{f})\,, \quad b_{1}={{2}\over{3}}\,
   (153-19\,n_{f})\,,
   \quad t=\log (Q^{2}/\mu^{2})\,. \eq
   One obtains from Eq.(17)\,\,\,($Z=1/{\alpha_s})$:
   \bq Z(\mu^{2})=Z(Q^2)-{{b_o}\over{4\,\pi}}\,\log (Q^2/\mu^{2})-
   \Delta\,\log
   \left({{Z(Q^2)+\Delta}\over{Z(\mu^{2})+\Delta}}\right)\,. \eq
   We use as an input the value:
   \bq \alpha_{s}(M^{2}_{W})=\,0.118\,, \eq
   which corresponds to: $\Lambda^{(5)}_{\overline {MS}}\se 200\,MeV,\,\,
   \Lambda^{(4)}_{\overline {MS}}\se 300\,MeV\,.$ One obtains then from\\
   Eqs.(19),(20)
   \footnote{The quark masses are found below.}:
   \bq  \alpha_{s}(\mu=M_{b}=5.04\,GeV)\se \, 0.204\,,\quad
     \alpha_{s}(\mu=M_{c}=1.65\,GeV)\se \, 0.310\,,\eq
    and the coefficients $C_i$ are equal to:
   \bq C_{-}(M_c)\se 1.770\,,\quad C_{+}(M_c)\se 0.762\,, \quad C_{1}(M_c)
     \se 1.266\,, \quad C_{2}(M_c)\se -0.504\,.\eq

   The decay width of a hadron containing a heavy quark:
   \bq \Gamma={{1}\over{2\,M_H}}\,\int d\,x\,<M_H|\,H_{W}(x)\, H_{W}(0)\,|
   M_H> \equiv {{1}\over{2\,M_H}}\,<H|\,L_{eff}(0)\,|H>\,, \eq
   can be represented in the form of the operator expansion \cite{BUV1}:
   \bq L_{eff}= \sum C_{i}\,O_{i}(0)\,, \eq
   where the leading term is the dimension 3 operator ${\bar Q}Q$ which
   describes (at $M_Q\ra \infty$) the free quark decay,\, fig.1. The next
   term is the dimension 5 operator ${\bar Q}\sigma G Q$ which describes
   an emission of a soft gluon,\, fig.2,\, etc. \\

   \begin{center} \{Fig.1\} \end{center}

  \begin{center} {\bf Fig.\,1}\,\,\,\, The Born contribution \end{center}

  \begin{center} \{Fig.2\} \end{center}

  \begin{center} {\bf Fig.\,2}\,\,\,\, The soft gluon emission giving rise
  the correction $O(1/M_Q^2)$ \end{center}

   If we confine ourselves temporarily to these terms only,\, then all
   $D^{o,\pm,s}$-mesons will have equal decay widths,\, and
   (on account of the radiative correction \cite{A1})
   it can be represented in the form \cite{BUV1}:
   \bq \Gamma_{nl}^{o}\se \Gamma_{o}z_{o}\left({{2\,C_{+}^{2}
   +C_{-}^2}\over{3}}\right){{<D|\,{\bar c}\,c\,|D>}\over{2\,M_D}}
   \left [ I_{rad}\,\left ( 1-2\,{{z_1}\over{z_o}}\Delta_{G}\right )+
   4\frac{z_2}{z_o}\left ({{C_{-}^{2}-C_{+}^2}\over
   {2\,C_{+}^{2}+C_{-}^2}}\right)\Delta_{G}\right ].\eq
   Here:
   \bq I_{rad}=\left ({{2\,C_{+}^2}\over{2\,C_{+}^{2}+C_{-}^2}}\right )\,
   \left [ 1-{{2}\over{3}}\,\pi\,\alpha_{s}(M_{c}^{2})+{{43}\over{12}}\,
   {{\alpha_{s}(M_{c}^2)}\over{\pi}}\right ]+ & \nonumber \\
   & \displaystyle \left ({{C_{-}^2}\over{2\,C_{+}^{2}+C_{-}^2}}\right )\,
   \left [ 1-{{2}\over{3}}\,\pi\,\alpha_{s}(M_{c}^{2})+{{25}\over{3}}\,
   {{\alpha_{s}(M_{c}^2)}\over{\pi}}\right ]\,, \eq
   \bq z_{o}=(1-8x+8x^3-x^4-12\,x^2\,\log x)\,,\,
   z_{1}=(1-x)^4\,,\, z_{2}=(1-x)^3\,,\, x=\frac{m_s^2}{M_c^2},   \eq
   \bq \Gamma_{o}={{G_{F}^{2}\,M_{c}^{5}}\over{64\,\pi^3}}\,, \quad
   {{<D|\,{\bar c}\,c\,|D>}\over{2\,M_D}}\se \left (1+{{1}\over{2}}\,
   \Delta_{G}-\Delta_{K} \right )\,, \eq
   \bq \Delta_{G}=\frac{1}{M_{c}^2}\,{{<D|\,{\bar c}\,(M_c^2-p_{c}^{2})
   \,c\,|D>}\over{<D|\,{\bar c}\,c\,|D>}}={{1}\over{M_{c}^2}}\,
   {{<D|\,{\bar c}\,
   {{i}\over{2}}\,g_{s}\,\sigma_{\mu\,\nu}\,G_{\mu\,\nu}^{a}\,
   {{\lambda^a}\over{2}}\,c\,|D>}\over{<D|\,{\bar c}\,c\,|D>}}\,, \eq
   \bq \Delta_{K}={{1}\over{2\,M_{c}^2}}\,{{<D|\,{\bar c}\,({\vec{p}}^{2})
     \,c\,|D>}\over{<D|\,{\bar c}\,c\,|D>}}
     \equiv {{<({\vec{p}}^{2})_{c}>}\over{2\,M_{c}^2}}\,, \eq
   where $p_c$ is the c-quark 4-momentum operator,\, $\vec p$ is its 3-
   momentum operator (in the D-meson rest frame). Let us recall that the
   radiative corrections in Eq.(26) are calculated at $m_s=0$.

   The lifetime differences appear first on account of the 4-quark operators,\,
   and we present here the contributions to $\delta L^{(c)}_{eff}$ from figs.
   3\,a,\,b,\,c-diagrams (see also Appendix):
   \bq \delta L^{c}_{eff}(M_c)={{G_F^2\,|V_{cs}|^2}\over{2\,\pi}}\,\left
   \{\,g_{\mu\,\nu}\,{\bar \lambda}^2\,L^{d}_{\mu\,\nu}+T_{\mu\,\nu}\left (
   \,L^{u}_{\mu\,\nu}+L^{s}_{\mu\,\nu}\,\right )
   \right \}_{\mu=M_{c}}\,+\cdots, \eq
   \bq L^{d}_{\mu\,\nu}=\left \{\,S_d^o\,
   (\,{\bar c}\,\Gamma_{\mu}\,d\,)\,({\,\bar d}\,\Gamma_{\nu}\,c\,)+
   O_d^o\,({\,\bar c}\,\Gamma_{\mu}{{\lambda^a}\over{2}}\,d\,)\,(\,{\bar d}\,
   \Gamma_{\nu}{{\lambda^a}\over{2}}\,c\,)\,\right \}\,, \eq
   \bq L^{u}_{\mu\,\nu}=\left \{\,S_u^o\,
   (\,{\bar c}\,\Gamma_{\mu}\,u\,)\,(\,{\bar u}\,\Gamma_{\nu}\,c\,)
   +O_u^o\,(\,{\bar c}\,
   \Gamma_{\mu}{{\lambda^a}\over{2}}\,u\,)\,(\,{\bar u}\,
   \Gamma_{\nu}{{\lambda^a}\over{2}}\,c\,)\,\right \}\,, \eq
   \bq L^{s}_{\mu\,\nu}=\left \{\,S_s^o\,
   (\,{\bar c}\,\Gamma_{\mu}\,s\,)\,(\,{\bar s}\,\Gamma_{\nu}\,c\,)
   +O_s^o\,(\,{\bar c}\,
   \Gamma_{\mu}{{\lambda^a}\over{2}}\,s\,)\,(\,{\bar s}\,
   \Gamma_{\nu}{{\lambda^a}\over{2}}\,c\,)\,\right \}\,. \eq
   \bq T_{\mu\,\nu}={{1}\over{3}}\left ( \lambda_\mu\,\lambda_\nu-\lambda^2\,
   g_{\mu\,\nu}\right )\,,\eq
   where $\lambda$ (or ${\bar \lambda}\,$) is the total 4-momentum of the
   integrated quark pair. It can be read off from each diagram
   and differ from $p_c$ by the spectator quark momenta.\\

   \begin{center} \{Figs.\,3a,\,3b,\,3c\} \end{center}

   \begin{center} {\bf Fig.\,3}\,\,\,\, The diagrams contributing into
   the 4-fermion operators \end{center}

  The coefficients $S_i^o$ and $O_i^o$ are:
   \bq S_d^o={{1}\over{3}}\,(C_{1}^{2}+C_{2}^{2})+2\,C_{1}\,C_{2}\,,\quad
   O_d^o=2\,(C_{1}^{2}+C_{2}^{2})\,,\eq
   \bq S_u^o=3\,(C_{2}+{{1}\over{3}}\,C_{1})^2\,,
   \quad O_u^o=2\,C_{1}^{2}\,,\quad
   S_s^o=3\,(C_{1}+{{1}\over{3}}\,C_{2})^2\,, \quad O_s^o=2\,C_{2}^{2}\,, \eq
   while $C_1,\, C_2$ are given in Eq.(22). Therefore:
   \begin{displaymath}
   {\left ( \begin{array}{cc}
   S_d^o & O_d^o \\ S_u^o & O_u^o \\ S_s^o & O_s^o
   \end{array}  \right  )}_{\mu=M_c}\se
   {\left ( \begin{array}{cc}
   -0.66 & 3.71 \\ 0.02 & 3.20 \\ 3.61 & 0.51
   \end{array}  \right  )}_{.}
   \end{displaymath}
   The leading contribution at $M_c\ra \infty ( \lambda \se{\bar \lambda}
   \se p_c \se M_c)$ in Eqs.(31)-(34) coincides with those
   obtained before in \cite{VS0},\,\cite{BGT},\,\cite{VS}.
   As will be shown below,\, however,\, it is
   important for calculations with charmed mesons to account for
   nonzero momenta of the spectator quarks\,: $\lambda \neq {\bar
   \lambda} \neq M_c$\,.

   The operators entering $\delta L^{(c)}_{eff}$ are normalized at the point
   $\mu=M_{c}$. So, it is reasonable to separate out explicitly the "hybrid"
   log effects \cite{VS},\cite{VSG},\cite{PW} by renormalizing
   to the point $\mu=\mu_o$,\, before
   the calculation of matrix elements. The 4-quark operators in
   $\delta L_{eff}$ are renormalized as\, \cite{VSG}
   (we neglect possible changes in the renormalization formulae
   originating from the presense of the spectator quark momentum operators in
   ${\bar \lambda}$) :
   \bq (S_{\mu\,\nu})_{\mu=M_Q}=\left [\,(\tau_{co}-{{\tau_{co}-1}\over{9}})\,
   S_{\mu\,\nu}-{{2}\over{3}}\,(\tau_{co}-1)\,O_{\mu\,\nu}\,\right ]_
   {\mu=\mu_o}\,, \eq
   \bq (O_{\mu\,\nu})_{\mu=M_Q}=\left [
   (1+{{\tau_{co}-1}\over{9}})\,
   O_{\mu\,\nu}-{{4}\over{27}}\,(\tau_{co}-1)\,S_{\mu\,\nu}\,\right ]_
   {\mu=\mu_o}\,, \eq
   \bq S^q_{\mu\,\nu}=(\,{\bar c}\,\Gamma_{\mu}\,q\,)\,(\,{\bar q}\,\Gamma_
   {\nu}\,c\,)\,, \quad
   O^q_{\mu\,\nu}=(\,{\bar c}\,\Gamma_{\mu}{{\lambda^a}\over{2}}\,q\,)\,
   (\,{\bar q}\,\Gamma_\nu {{\lambda^a}\over{2}}\,c\,)\,, \eq
   \bq \tau_{co}=\left ({{\alpha_{s}(\mu_o^2)}
    \over{\alpha_{s}(M_c^2)}}\right )^{9/2b{_o}}\,.\eq

   As for a concrete value of $\mu_o$,\, let us point out that the
   mean value of the vacuum quark 4-momentum squared is \cite{Ioffe}:
   \bq <-k^2>_o\equiv {{<0|\,{\bar q}\,D_\mu^2\,q\,|0>}\over{<0|\,{\bar q}\,
   q\,|0>}}={{<0|\,{\bar q}{{i}\over{2}}\,g_s\,\sigma_{\mu\,\nu}G_{\mu\,\nu}
   ^a{{\lambda^a}\over{2}}\,q\,|0>}\over{<0|\,{\bar q}\,q\,|0>}}
   \se 0.4\,GeV^2\,.\eq
   This number determines the characteristic scale of nonperturbative
   interactions,\, so that the scale $\mu_o$ must not be chosen below
   of this value. In what follows we will use the value:
   \bq \mu_o^2=0.5\,GeV^2\,,\quad \alpha_s(\mu_o^2=0.5\,GeV^2)=0.58\,.\eq
   \bq \eta_{co}={{\alpha_{s}
   (\,\mu_{o}^{2}\,)}\over{\alpha_{s}(\,M_{c}^{2}\,)}}\,\se 1.87\,,\quad
   \tau_{co}=\eta_{co}^{1/2}\se 1.37\,.\eq
   Now,\, after renormalization $\delta L^{(c)}_{eff}$ has the form:
   \bq \delta L^{(c)}_{eff}(\mu_{o}\equiv \Delta L(\mu_o)+
   L^{(c)}_{PNV}(\mu_o)\,,\eq
   \bq \Delta L(\mu_{o})={{G_{F}^2}\over{2\,\pi}}\,\left \{\,
   T_{\mu\,\nu}\,L^{u}_{\mu\,\nu}+\Lambda^d+\Lambda^s\,
   \right \}_{\mu_{o}}\,,\eq
   where:
   \bq \Lambda^d=\left [\,|V_{ud}|^2\,g_{\mu\,\nu}\,{\bar \lambda}^2
   \,L_{\mu\,\nu}^d+|V_{cd}|^2\,T_{\mu\,\nu}\left (\,S_s\,S_{\mu\,\nu}^d+
   O_s\,O_{\mu\,\nu}^d\,\right) \right]\,,\eq
   \bq \Lambda^s=\left [\,|V_{cs}|^2\,T_{\mu\,\nu}\,L_{\mu\,\nu}^s+
   |V_{us}|^2\,g_{\mu\,\nu}\,{\bar \lambda}^2\,\left (\,S_d\,S_{\mu\,\nu}
   ^s+O_d\,O_{\mu\,\nu}^s\,\right ) \right ]\,, \eq
   and the coefficients are:
   \begin{displaymath}
   {\left ( \begin{array}{cc}
   S_d & O_d \\ S_u & O_u \\ S_s & O_s
   \end{array}  \right )}_{\mu=\mu_o}\se \tau_{co}{\left ( \begin{array}{cc}
   -0.73 & 2.93 \\- 0.11 & 2.43 \\ 3.48 & -0.26
   \end{array} \right )}\se {\left (\begin{array}{cc}
   -1.07 & 4.02 \\ -0.15 & 3.33 \\ 4.76 & -0.36
   \end{array} \right )}_{.}
   \end{displaymath}

   The "penguin nonvalence" (PNV) term,\, $ L^{(c)}_{PNV}$,\,
   originates from the contribution of the diagram in fig.4.

   \begin{center} \{Fig.\,4\} \end{center}

   \begin{center} {\bf Fig.\,4}\,\,\,\, The nonvalence factorizable
   penguin contribution (PNV) \end{center}

   With logarithmic accuracy $L_{PNV}$  can be
   easily obtained from Eqs.(31)-(34) by extracting the term $\sim D_{\mu}
   G_{\mu\,\nu}=-g_{s}\,J_{\nu}$ from the light quark operator $[q{\bar q}]$.
   For instance:
   \bq [\,d\,\bar d\,] \ra -{{1}\over{2}}\,
   (\,\gamma_{\rho}{{\lambda^a}\over{2}}
   \,)\,\left [\,{\bar d}\,\gamma_{\rho}{{\lambda^a}\over{2}}\,d \right ] \ra
   - {{1}\over{6}}\,(\,\gamma_{\rho}{{\lambda^c}\over{2}}\,)\,J^a_{\rho}(3) \ra
   {{1-\eta_{co}^{-2/b_o}}\over{6}}\,(\,\gamma_{\rho}{{\lambda^c}\over{2}}\,)\,
   J^a_{\rho}(3)\,, \eq
   \bq J^a_{\rho}(3)=\left (\,{\bar u}\,\gamma_{\rho}{{\lambda^c}\over{2}}\,u+
   (\,d\,)+(\,s\,) \right )\,, \eq
   where the last step in Eq.(49) describes the evolution of $J^a_{\rho}$
   from $M_c$ to $\mu_o$.
   Proceeding in this way (and using $\lambda\se {\bar \lambda}\se p_c,\,\,
   \lambda^2\se {\bar \lambda}^2\se M^2_c,\,$ which is justified
   for these contributions,\, see fig.4) we obtain:
   \bq L^{(c)}_{PNV}(\mu_o) \se {{G_{F}^{2}}\over{2\,\pi}}
   (1-\eta_{co}^{-2/9})\,
   M_c^2\, J^a_\rho(3)\,\left [\,\,{\bar c}\,\gamma_\rho
   {{\lambda^a}\over{2}}\,(\,N_v+N_a\,\gamma_5\,)\,c\,\right ]_{\mu_o}\,, \eq
   \bq N_{v}={{1}\over{3}}\left (\,A_u+A_s-2\,A_d\, \right)\se 2.54\,,\eq
   \bq N_{a}={{1}\over{3}}\left (\,\frac{1}{3}\,A_u+\frac{1}{3}\,A_s-2\,
   A_d\,\right )\se 1.62\,,\eq
   \bq A_{i}=(\,S_{i}-{{1}\over{6}}\,O_{i}\,)\,. \eq

   For the semileptonic decays,\, $\delta L^{lept}_{eff}$ can be easily
   obtained from Eqs.(45)-(48),\,(51)) by corresponding replacements
   and has the form :
   \bq \delta L^{lept}_{eff}\equiv \Delta L^{lept}+L_{PNV}^{lept}\,, \eq
   \bq \Delta L^{lept}\se {{G_F^2}\over{2\,\pi}}\,
   T_{\mu\,\nu} \left [\,|V_{cs}|^2\, \left (\,
   1.327\,\,{\bar c}\,\Gamma_\mu\,s \cdot {\bar s}\,\Gamma_
   \nu\,c- 0.245\,\,{\bar c}\,
   \Gamma_\mu {{\lambda^a}\over{2}}\,s\cdot {\bar s}\,
   \Gamma_\nu {{\lambda^a}\over{2}}\,c\, \right )+
   \right. & \nonumber \\
   & \displaystyle \left. |V_{cd}|^2\, \left (\,
   1.327\,\,{\bar c}\,\Gamma_\mu\,d \cdot {\bar d}\,\Gamma_
   \nu\,c- 0.245\,\,{\bar c}\,
   \Gamma_\mu {{\lambda^a}\over{2}}\,d\cdot {\bar d}\,
   \Gamma_\nu {{\lambda^a}\over{2}}\,c\, \right )
   \, \right ]_{\mu_o}\,,\eq
   \bq L_{PNV}^{lept}\se {{G_F^2}\over{2\,\pi}}{{\left (\,1-\eta_{co}^{-2/9}
   \,\right )}\over{3}}\,\tau_{co}\,M^2_c\left [\,J^a_{\rho}(3)\cdot{\bar c}
    \,\gamma_{\rho}\left (1+{{1}\over{3}}\gamma_5\right )
   {{\lambda^a}\over{2}}\,c\, \right ]_{\mu_o}.\eq \\

  {\bf 3b. \quad  General formulae:\, b-quark}\\

   The relevant part of the weak Hamiltonian has the form:
\bq H_{W}^{b}(\mu=M_b)=\frac{G_F}{\sqrt 2}\,V_{c b}\,V^{*}_{u d}
\left\{\,C_{1}(M_b)\, {\bar c}\,\Gamma_\mu b \cdot {\bar d}\,\Gamma_\mu u+
C_{2}(M_b)\,{\bar d}\,\Gamma_\mu b \cdot
{\bar c}\,\Gamma_\mu u\, \right \}\,+\cdots \eq
\bq \alpha_s(\mu=M_{b}=5.04\,GeV)\se \, 0.204\,,\eq
\bq C_{-}(M_b)\se 1.385\,,\quad C_{+}(M_b)\se 0.855\, \quad C_{1}(M_b)
\se 1.120\,, \quad C_{2}(M_b)\se -0.265\,.\eq

For the beauty-hadrons one can obtain $\delta L^{b}_{eff}$
from $\delta L^{c}_{eff}$ by evident replacements:\\
\bq \delta L^{b}_{eff}(M_b)=\frac {G_F^2\,|V_{cb}|^2}{2\,\pi}\,
\left \{\,\left ( {\bar L}_{u}+ L_d+L_c\,\right )+
\left ( {\bar L}_c+ L_s+{\widetilde L}_c\,\right )
\right \}_{\mu=M_b}\,+\cdots\,, \eq
\bq {\bar L}_{u}=g_{\mu\,\nu}\,{\bar \lambda}^{2}\,\left (1-\frac
{M_c^2}{\bar \lambda^2}\right )^2 \left \{\,S_d^o\,
(\,{\bar b}\,\Gamma_{\mu}\,u\,)\,({\,\bar u}\,\Gamma_{\nu}\,b\,)+
O_d^o\,({\,\bar b}\,
\Gamma_{\mu}\frac{\lambda^a}{2}\,u\,)\,(\,{\bar u}\,
\Gamma_{\nu}\frac{\lambda^a}{2}\,b\,)\,\right \}\,, \eq
\bq {\bar L}_{c}=g_{\mu\,\nu}\,{\bar \lambda}^{2}\,\left (1-\frac
{M_c^2}{\bar \lambda^2}\right )^2 \left \{\,S_d^o\,
(\,{\bar b}\,\Gamma_{\mu}\,c\,)\,({\,\bar c}\,\Gamma_{\nu}\,b\,)+
O_d^o\,({\,\bar b}\,
\Gamma_{\mu}\frac{\lambda^a}{2}\,c\,)\,(\,{\bar c}\,
\Gamma_{\nu}\frac{\lambda^a}{2}\,b\,)\,\right \}\,, \eq
\bq L_{d}=T^{(c)}_{\mu\,\nu}(\lambda) \left \{\,S_u^o\,
(\,{\bar b}\,\Gamma_{\mu}\,d\,)\,(\,{\bar d}\,\Gamma_{\nu}\,b\,)
+O_u^o\,(\,{\bar b}\,\Gamma_{\mu}\frac{\lambda^a}{2}\,d\,)\,(\,{\bar d}\,
\Gamma_{\nu}\frac{\lambda^a}{2}\,b\,)\,\right \}\,, \eq
\bq L_{s}=\,T^{({\bar c}c)}_{\mu\,\nu}(\lambda) \left \{\,S_u^o\
(\,{\bar b}\,\Gamma_{\mu}\,s\,)\,(\,{\bar s}\,\Gamma_{\nu}\,b\,)
+O_u^o\,(\,{\bar b}\,\Gamma_{\mu}\frac{\lambda^a}{2}\,s\,)\,(\,{\bar s}\,
\Gamma_{\nu}\frac{\lambda^a}{2}\,b\,)\,\right \}\,, \eq
\bq L_{c}=T_{\mu\,\nu}(\lambda) \left \{\,S_s^o\,
(\,{\bar b}\,\Gamma_{\mu}\,c\,)\,(\,{\bar u}\,\Gamma_{\nu}\,c\,)
+O_s^o\,(\,{\bar b}\,
\Gamma_{\mu}\frac{\lambda^a}{2}\,c\,)\,(\,{\bar c}\,
\Gamma_{\nu}\frac{\lambda^a}{2}\,b\,)\,\right \}\,, \eq
\bq {\widetilde L}_{c}=\,T^{(c)}_{\mu\,\nu}(\lambda)\left \{\,S_
s^o\,(\,{\bar b}\,\Gamma_{\mu}\,c\,)\,(\,{\bar c}\,\Gamma_{\nu}\,b\,)
+O_s^o\,(\,{\bar b}\,\Gamma_{\mu}\frac{\lambda^a}{2}\,c\,)\,(\,{\bar c}\,
\Gamma_{\nu}\frac{\lambda^a}{2}\,b\,)\,\right \}\,, \eq
where the dots in the above formulae denote the Cabibbo-suppressed
contributions.
Here $(x=M_c^2/{\lambda^2})$\,:
\bq T_{\mu\,\nu}(\lambda)=\frac{1}{3} \left (\lambda_\mu\,
\lambda_\nu-{\lambda^2}\,g_{\mu\,\nu}\right )\,,\eq
\bq T^{(c)}_{\mu\,\nu}(\lambda)=\frac{1}{3}(1-x)^2 \left [ \left (\lambda_\mu\,
\lambda_\nu-{\lambda^2}\,g_{\mu\,\nu}\right )(1+2\,x) +\frac{3}{2}\,
g_{\mu\,\nu}\,{\lambda^2}\,x \right ]\,,\eq
\bq T^{({\bar c}c)}_{\mu\,\nu}(\lambda)=\frac{1}{3}\sqrt{1-4\,x}\left [(1+2\,x)
\,\lambda_\mu\,\lambda_\nu-(1-x)\,{\lambda^2}\,g_{\mu\,\nu}\right ]\,.\eq

The coefficients in Eqs.(61-(67)):
\begin{displaymath}
{\left ( \begin{array}{cc}
S_d^o & O_d^o \\ S_u^o & O_u^o \\ S_s^o & O_s^o
\end{array} \right )}_{\mu=M_b}\se {\left ( \begin{array}{cc}
-0.15 & 2.65 \\0.03 & 2.51 \\ 3.20 & 0.14
\end{array} \right )}_{,}
\end{displaymath}
become after the renormalization,\, $M_b\ra \mu_{o}$:\\
i) for the terms $L_d,\,L_s,\,{\bar L}_u$,\, for which the renormalization
factor is
\bq \tau_{bo}=\eta_{bc}^{27/50}\tau_{co}\se 1.715\,, \quad
\eta_{bc}=\frac{\alpha_{s}(M_c^2)}{\alpha_{s}(M_b^2)}\se 1.52\,,\eq
\begin{displaymath}
{\left ( \begin{array}{cc}
S_d & O_d \\ S_u & O_u \\ S_s & O_s
\end{array} \right )}_{\mu=\mu_o}\se
\tau_{bo}{\left ( \begin{array}{cc}
-0.31 & 1.71 \\ -0.12 & 1.57 \\ 3.04 & -0.80
\end{array} \right )}\se
{\left ( \begin{array}{cc}
-0.53 & 2.93 \\-0.21 & 2.69 \\ 5.21 & -1.37
\end{array} \right )}_{,}
\end{displaymath}
ii) for the terms $L_c,\,{\bar L}_c,\,{\widetilde L}_c$,\, for which the
renormalization stops at $\mu=M_c$ and the renormalization factor is:
\bq \tau_{bc}= \eta_{bc}^{27/50}\se 1.253: \eq
\begin{displaymath}
{\left ( \begin{array}{cc}
S_d & O_d  \\ S_s & O_s
\end{array} \right )}_{\mu=M_c}\se \tau_{bc}
{\left ( \begin{array}{cc}
-0.23 & 2.20 \\ 3.12 & -0.32
\end{array} \right )}\se
{\left ( \begin{array}{cc}
-0.29 & 2.75 \\ 3.91 & -0.40
\end{array} \right )}_{.}
\end{displaymath}

The corresponding term $ L^{(b)}_{PNV}$ has the form:
\bq L^{(b)}_{PNV}\se \frac{G_F^2\,|V_{cb}|^2}{2\,\pi}\,M_b^2 \left \{
J^a_{\rho}(3)\left [\,{\bar b}\frac{\lambda^a}{2}\gamma_{\rho}\left (A+B
\gamma_5 \right )b \right ]+
J^a_{\rho}(c)\left [{\bar b}\frac{\lambda^a}{2}\gamma_{\rho}\left (
C+D\gamma_5\right ) b\right ] \right \}, \eq
\bq A=\lambda_1\left(A_s^o-A_d^o\right )+\lambda_2\left(A_u^o-
A_d^o\right )\se 0.40;\,\eq
\bq B=\lambda_1\left(\frac{1}{3}A_s^o-A_d^o\right )+\lambda_2\left
(\frac{1}{3}A_u^o-A_d^o\right )\se 0.27,\eq
\bq C=\lambda_3 \left ( A_s^o+A_u^o-2\,A_d^o \right )\se 0.31\,;\,\,
D=\lambda_3 \left (\frac{1}{3}A_s^o+\frac{1}{3}A_u^o-2\,A_d^o
\right )\se 0.17\,,\eq
\bq \lambda_3=\frac{1-\eta_{bc}^{-8/25}}{2} \tau_{bc}\,,\,\,
\lambda_1=\lambda_3\,\eta_{co}^{-2/9}\,\tau_{co}\,,\,\,
\lambda_2=\frac{2}{3}\left (1-\frac{1+3\eta_{bc}^{-8/25}}{4}\eta_{co}^
{-2/9}\right )\tau_{bo}\,,\eq
\bq A_i^o=\left (S_i^o-\frac{1}{6} O_i^o\,\right ),\quad
J^a_{\rho}(3)={\bar u}\frac{\lambda^a}{2}\gamma_{\rho}\,u+(d)+(s)\,,\,\,
J^a_{\rho}(c)={\bar c}\frac{\lambda^a}{2}\gamma_{\rho}\,c\,.\eq

For the leptonic decays,\, $B\ra l+\nu+X,\,\,\, l=e,\,\mu,\,$ we have:
\bq \delta L^{lept}_{eff}=\Delta L^{lept}+L^{lept}_{PNV}\,,\eq
\bq \Delta L^{lept}\se \frac{G_F^2\,|V_{cb}|^2}{2\,\pi}
T_{\mu\,\nu}(\lambda)\left [\,
1.225\,{\bar b}\,\Gamma_\mu c \cdot {\bar c}\,\Gamma_
\nu b- 0.169\,{\bar b}\,
\Gamma_\mu \frac{\lambda^a}{2} c\cdot {\bar c}\,
\Gamma_\nu \frac{\lambda^a}{2} b\, \right ]_{\mu_o},\eq
\bq L_{PNV}^{lept}\se \frac{G_F^2\,|V_{cb}|^2}{2\,\pi}\,
\frac{\left (1-\eta_{bc}^{-8/25}\right )}
{4}\,\tau_{bo}\,\eta_{co}^{-2/9}\,M^2_b \times & \nonumber \\
& \displaystyle \left [\, {\bar b}\,\gamma_{\rho}
\left (1+\frac{1}{3}\gamma_5\right )\frac{\lambda^a}{2}\, b\, \right ]\cdot
\left [\,J^a_{\rho}(3)+ \tau_{co}^{-1}\,\eta_{co}^{2/9}\,J^a_{\rho}(c)\,
\right ]_{\mu_o}\se & \nonumber \\
& \displaystyle  \frac{G_F^2\,|V_{cb}|^2}{2\,\pi}\,
0.046\,M^2_b\,\left [\,{\bar b}\,\gamma_{\rho} \left
(1+\frac{1}{3}\gamma_5\right )\frac{\lambda^a}{2}\,b\,\right ]\cdot \Bigl
[ J^a_{\rho}(3)+0.85\,J^a_{\rho}(c)\Bigr ]_{\mu_o}\,.\eq \\

 {\bf 4.\quad  $D^{+}\ra l\nu+X$. \quad Determination of $M_c$.} \\

 The purpose of this section is to find out the value of the c-quark
"hard pole mass" (see sect.2),\,
 \,$M_c$,\, using the experimental data for the semileptonic width
 \cite{PDG},\,\cite{MD}:
 \bq \Gamma_{sl}(D^{+}\ra l\nu+X)=(1.06\pm 0.11)\cdot 10^{-13}\,GeV\,.\eq
 The expression for $\Gamma_{sl}$ can be represented in the form (see
 \cite{BUV1} and sect.3a):
 \bq \Gamma_{sl}(\,D^{+}\,)\se \Gamma_{sl}^{o}\, \left \{\,z_{o}\,
   I_{rad}\,{{<D|\,{\bar c}\,c\,|D>}\over{2\,M_D}}\,
   \left [ 1-2\,{{z_1}\over{z_o}}\,\Delta_{G}\right ]
   +\delta_{lept}^{(c)}\,\right \}\,, \eq
   \bq \Gamma_{sl}^{o}={{G_{F}^{2}\,M_{c}^5}\over{192\,\pi^3}}\,,\quad
   I_{rad}\se \left [1-{{2}\over{3}}\,{{\alpha_{s}(M_c^2)}\over{\pi}}\,
   f_{o}({{m_{s}^2}\over{M_{c}^2}})\,\right ]\,,\eq
   where $\delta_{lept}^{(c)}$ in Eq.(83) is the contribution of $\delta
   L^{lept}_{eff}$ into $\Gamma_{sl}$,\, and the explicit form of $f_o
   (x)$ can be found in \cite{Beh},\,\cite{CM}.
   We use below \footnote
   {From our viewpoint,\, the value: $<{\vec p}^2>\se 0.6\,GeV^2$ available in
   the literature \cite{PBB} is overestimated. Let us recall (see Eq.(42)) that
   the mean value of the vacuum quark 4-momentum squared
   is: $<-k_{\mu}^2>_o=4/3\,<{\vec k}^2>_o\se 0.4\,GeV^2,\,$ and the quarks
   inside the pion have their momenta on the average somewhat less than
   those of the vacuum quarks \cite{CZ5}.
   Let us point out also that $<{\vec p}^2>$ enters here to
   $<D|{\bar c}\,c|D>$ only and plays no essential role.}:
   \bq m_{s}\se 150\,MeV\,, \,\, <(\,{\vec p}^{2}\,)_{c}>\se 0.3\,GeV^2\,,
   \quad \Delta_{G}\se {{3}\over{4}}\,
   {{M_{D^{*}}^{2}-M_{D}^2}\over{M_{c}^2}}\,.\eq
   The value of $M_c$ can be determined now from a comparison
   of Eqs.(82) and (83).
   Because the dependence of $\Gamma_{sl}$ on $M_c$ is highly nonlinear,\, it
   is more convenient to proceed in an opposite way. Namely,\, let us show
   that Eq.(83) reproduces the experimental value at $M_c\se 1.65
   \,GeV$. We have:
   \bq \alpha_{s}(\,M_{c}^{2}\,)\se 0.310\,, \quad z_{o}\se 0.938\,,\quad
   z_{1}\se 0.967\,, \eq
   \bq \Delta_{G}\se 0.15\,,\quad \Delta_{K}\se 0.055\,,
   \quad f_{o}\se 3.25\,,\eq
   \bq \Gamma_{sl}^{o}\se 2.80\cdot 10^{-13}\,GeV\,,\eq
   \bq I_{rad}\se 0.786\,, \quad {{<D|\,{\bar c}\,c\,|D>}\over{2\,M_D}}
   \se 1.02\,,\quad \left [ 1-2\,{{z_1}\over{z_o}}\,
   \Delta_{G}\right ] \se 0.691\,. \eq

   Substituting all this into Eq.(83) one obtains:
   \bq \Gamma_{sl}(\,D^{+}\,)\se \Gamma_{sl}^{o}\,(\,0.52+\delta^{(c)}
   _{lept}\,)\,.\eq
   The quantity $\delta_{lept}^{(c)}$ is calculated in sect.11 and is:
   \bq \delta^{(c)}_{lept}\se -0.13\,.\eq
   So\,:
   \bq \Gamma_{sl}(D^+)\se 0.39\,\Gamma_{sl}^{o}\se 1.09
   \cdot 10^{-13}\,GeV\,, \eq
   in agreement with Eq.(82).\\

   {\bf 5. \quad  Mass formulae. Determination of $M_b$ and ${\bar \Lambda}$}\\

   To find out the value of the b-quark "hard pole" mass,\, $M_b$,\,
   we use the mass formula which has the form (see,\, for instance,\,
   \cite{Neu} and sect.2) :
   \bq M_{D}= M_{c}+{\bar \Lambda}+\frac{1}{M_c}\delta m^2+
   O\left (\frac{\Lambda_{QCD}^3}{M_c^2}\right ),\eq
   \bq \delta m^2\equiv \frac{1}{2}\,\frac{<D(p)|\,
   {\bar c}\,\Delta\,c\,|D(p)>}{2\,M_D}\se
   \frac{1}{2}<\,{\vec p_c}^2\,>-\frac{3}{8}\,(\,M_{D^*}^2-M_D^2\,)\,,\eq
  \bq  \Delta=(i{\vec D})^2-\frac{i}{2}g_s\sigma_{\mu\,\nu}G_{\mu\,\nu}^a
   \frac{\lambda^a}{2}\,.\eq
   As it was pointed out in sect.2,\, the difference between the "hard
  pole mass",\, $M_c$,\, and the pole mass,\, $M_p$,\, can be absorbed by a
redefinition of $\Lambda_o$ appearing in the matrix element:
\bq \frac{1}{2\,M_D}\,<D|\,H_{light}\,|D>= \Lambda_o
 \left (\,1+O(\frac{\Lambda_{QCD}^2}{M_c^2})\,\right )\,,\eq
 or,\, equivalently,\, through the trace anomaly:
\bq \frac{1}{2\,M_D}<D|\,\frac{\beta (\alpha)}{2\pi\alpha}\,
G^2_{\mu\,\nu}\,|D>=\Lambda_o+\frac{2\,\delta {\tilde m}^2}{M_c}+O(
\frac{\Lambda_{QCD}^3}{M_c^2})\,,\eq
and by a redefinition of the next term,\, $(1/4 M_D)<D|\,{\bar c}\Delta\,
c\,|D>=\delta {\tilde m}^2\ra \delta m^2,\,$ etc.

Turning now to concrete numbers,\, we use  in Eq.(93):
   \bq M_{c}\se 1.65\, GeV\,,\quad  M_{D}\se 1.867\,GeV\,,& \nonumber \\
& \displaystyle  <\,{{\vec p}_c}^2\,>\se 0.3\,GeV^2\,,\quad
  (\,M_{D^{*}}^2-M_{D}^2\,)\se 0.543\,GeV^2\,, \eq
 and obtain:
   \bq 1.867\,GeV\se 1.65\,GeV+{\bar \Lambda}+91\,MeV-123\,MeV\,,\eq
   \bq {\bar \Lambda}\se 250\,MeV\,.\eq
   Using now this value of ${\bar \Lambda}$ and:
 \bq M_{B}\se M_{b}+{\bar \Lambda}+{{1}\over{2\,M_b}}<\,{\vec p_b}^2\,>-
   {{3}\over{8\,M_b}}\,(\,M_{B^{*}}^2-M_{B}^2\,)+
O(\frac{\Lambda_{QCD}^3}{M_b^2})\,.\eq
\bq M_B\se5.28\,GeV\,,\quad  <{{\vec p}_b}^2>\se 0.3\,GeV^2\,,\quad
 (M^{2}_{B^*}-M^{2}_{B})\se 0.488\,GeV^2\,,\eq
   one obtains from Eq.(101) \footnote {Really,\, it is clear that we
need not ${\bar \Lambda}$ at all to determine $M_b\,.$}:
   \bq M_b\se 5.04\,GeV\,.\eq \\

    {\bf 6\,.\quad  $B\ra l\,\nu+X\,$. Determination of $|V_{cb}|$\,.} \\

  The semileptonic decay width $"B\ra e\,\nu+X"\,$ is obtained from
  Eq.(83) by evident replacements. We have:
  \bq \Gamma_{sl}^{o}={{G_{F}^{2}\,M_{b}^{5}|V_{cb}|^2}\over
  {192\,\pi^3}}\,\se 1.19\cdot 10^{-13}\,GeV \left |{{V_{cb}}\over{0.040}}
  \right |^2\,,\eq
  \bq \alpha_{s}(\,M_{b}^{2}\,)\se 0.204\,, \quad z_{o}\se 0.460\,,\quad
  z_{1}\se 0.635\,, \quad f_{o}\se 2.46\,,\eq
  \bq {{<B|\,{\bar b}\,b\,|B>}\over{2\,M_B}}\se 1.00\,,\quad
  \left [ 1-2\,{{z_1}\over{z_o}}\,\Delta_{G}\right ] \se 0.96\,. \eq
  \bq I_{rad}\se \left [1-{{2}\over{3}}\,{{\alpha_{s}(M_b^2)}\over{\pi}}\,
  f_{o}({{M_c^2}\over{M_b^2}})\right ]\se 0.893\,.\eq
  So,\, (see Eq.(83):
  \bq \Gamma(B\ra e\,\nu+X)\se \Gamma_{sl}^{o}\left (\,0.394+\delta_{lept}^
  {(b)}\,\right )\,.\eq
  The quantity $\delta_{lept}^{(b)}$ is calculated in sect.11:
  \bq \delta_{lept}^{(b)}\se -\, 2\cdot 10^{-3}\,,\eq
  and is negligble. Therefore:
  \bq \Gamma(B\ra e\,\nu+X)\se 0.47\cdot 10^{-13}\,GeV\,\left |{{V_{cb}}
  \over{0.04}} \right |^2\,.\eq
  If we take,\, for instance,\, \cite{MD}
  \footnote{We use the LEP data for the b-quark lifetime and $Br (b\ra e\nu+X
  )$. When comparing these with the $\Upsilon (4S)$ data,\, one sees that the
  absolute value of $\Gamma (B\ra e\nu+X)\se \Gamma (b\ra e\nu+X)$ is the same,
\, while the LEP data give a smaller value for the total decay width. We
  prefer to use the LEP value for the following reasons. It gives the weighted
  average of the b-hadron widths. Because (see below) the B-meson widths are
  practically the same,\, and there is all the reason to expect that the
  b-baryon total widths are somewhat larger than those of the B-mesons (with
  the semileptonic widths being the same),\, the
  admixture of the b-baryons can only increase the weighted total decay width.
  So,\, the LEP data can be considered as giving the upper limit for the
  B-meson decay width and the lower limit for the semileptonic branching.} :
  \bq \tau (\,B\,)=(1.6\pm 0.04)\cdot 10^{-12}\,s\,, \quad \Gamma_{tot}(\,B\,)
  =(4.1\pm 0.1)\cdot 10^{-13}\,GeV\,,\eq
  \bq Br\, (B\ra l\,\nu+X)=(11.4\pm 0.5)\,\%\,,\eq
  then:
  \bq \Gamma\,(B\ra l\,\nu+X)=(0.47\pm 0.03)\cdot 10^{-13}\,GeV \left [
  \frac{Br(B\ra l\nu+X)}{11.4\%}\right ]\left [\frac{1.6\cdot 10^{-12}s}
  {\tau(B)}\right ],\eq
  and comparing with Eq.(110) we obtain:
  \bq |V_{cb}|\se 0.040 \left [ \frac{Br(B\ra l\nu+X)}{11.4\%}\right ]^{1/2}
  \left [\frac{1.6\cdot 10^{-12}s}{\tau(B)}\right ]^{1/2}.\eq \\

 {\bf 7.\quad  Calculation of $f_{D}\,,\,f_{B}$ }\\

  There is a large number of papers dealing with the calculation of the
  decay constants $f_{D}$ and $f_{B}$ with the help of the QCD sum rules.
  We point out here only early papers \cite{Sh},\,\cite{CZ1},\,\cite{AE},\,
  and the paper \cite{CZ2} close in spirit to this work,\, were it was proposed
  to use the correlators of chiral currents in the QCD sum calculations.

  Due to reasons which are explained below in detail (see also \cite{CZ2}),\,
 our approach here rests heavily on the use of the chiral currents correlators.
  One of the main difficulties which prevents the calculation of reliable
  results from the correlator of the pseudoscalar currents,\, is the very large
  radiative correction to the Born approximation for this correlator \cite
  {Br},\,\cite{BG1},\,\cite{BG2}.
  To avoide this difficulty,\, let us consider the following correlator of
  the chiral currents:
  \bq K_{1}(\,q^2\,)=i \int d\,x\,e^{i\,q\,x}<0|T \left \{ {\bar Q}
  (x)i(\,1+\gamma_{5}\,)\,q(x)\,, {\bar q}(0)
  i(\,1+\gamma_{5}\,)\,Q(0)\right \}|0>\equiv & \nonumber \\
  & \displaystyle  i\, \int d\,x\,e^{i\,q\,x}\,<0|T \left \{ P(x)\,P^{+}(0)-
  S(x)\,S^{+}(0)\,\right \}|0>\,, & \nonumber \\
  & \displaystyle  P(x)={\bar Q}i \gamma_{5}\,q(x)\,, \quad S(x)=
  {\bar Q}(x)\,q(x)\,. \eq

  Because $K_{1}(q^2)$ is the difference of the pseudoscalar and scalar current
  correlators,\, the pure perturbative contributions cancel completely in
  all orders of the perturbation theory (in the chiral limit). On the other
  hand,\, there appear additional scalar state contributions in the spectral
  density. Let us emphasize,\, however,\, that the mass differences between
  the lowest lying pseudoscalar and scalar resonances are sufficiently large
  nevertheless,\, both for the D and B mesons. So (after the Borel
  transformation),\, the scalar meson contributions are sufficiently
  suppressed in the sum rules.

  The diagrams giving the main contribution to $K_{1}(q^2)$ are shown in
  fig.5. The spectral density (for the D-meson) has the form:
  \bq \delta\, K(\,s\,)=r_{D}^{2}\,\delta (\,S-M_{D}^2\,)+ \cdots\,,\quad
  r_{D}=f_{D}\,M_{D}^2\,/\,M_{c}\,. \eq
   The sum rules obtained in a standard way have the form \footnote { We
  neglected in Eq.(118) the small contribution $\sim \,<{\bar q}\,q>^2$.
  The anomalous dimensions of the operators ${\bar q}D^{2}q$
  and ${\bar q}q$ are respectively: $(-2/3b_o)$ and $(4/b_o)\,.$ }:

  \bq r_{D}^2(\overline {M^2})=2\,M_{c}\,<0|q\,{\bar q}|0>_{\overline {M^2}}\,
  \Phi_{c}(M^2)\,,\eq
  \bq \Phi_{c}(M^2)=exp \left \{{{M_{D}^2-M_{c}^2}\over{M^2}}\right \}\,
  \left [\,1-\frac{m_{o}^2\,M_{c}^2}{4\,M^4}\left (\,1-2\,\frac{M^2}
  {M_c^2}\right )
  \right ]\,, \eq
  \bq \frac{m_o^2}{2}=\frac{<0|{\bar q}\frac{i}{2}\sigma_{\mu\,\nu}\,G_{\mu\,
  \nu}^{a}\,\frac{\lambda^a}{2}q|0>_{\overline {M^2}}}{<0|{\bar q}\,q\,
  |0>_{\overline {M^2}}}=
  \frac{<0|{\bar q}\,(\,i\,D_\mu\,)^2\,q\,|0>_{\overline {M^2}}}
  {<0|{\bar q}\,q\,|0>_{\overline {M^2}}}\,.\eq
  In Eqs.(117) and (119)\, ${\overline {M^2}}\se 1\,GeV^2\,$ (see below) is the
  normalization point of all operators and of the D-meson residue,\, $r_D\,$,
  and the number $m_o^2$ is also determined at this point. We will use for it
  the standard value determined previously \cite{Ioffe} at this scale:
  $m_o^2\se 0.8\,GeV^2$.

  \begin{center} \{Fig.\,5\} \end{center}

  \begin{center} {\bf Figs.\,5a,\,5b,\,5c} \,\,\,\, The diagrams contributing
  to the sum rule Eq.(118) \end{center}

  The power corrections due to the figs.5b,\,c diagrams do not exceed $\se
  35\%$ of the fig.5a contribution in the region $0.8\,GeV^2\le M^2 \le 1.5
  \,GeV^2\,$ and,\, on the other
  hand,\, the values of $M^2$ are sufficiently small for the contributions
  of higher states in the spectral density to be exponentially suppressed.
  The function $\Phi_{c}(M^2)$ varies only slightly in this region
  and the characteristic value of $M^2$ (at the extremum of $\Phi_{c}
  ( M^2)$\,) is:
  \bq \overline {M^2}\se 1.15\,GeV^2\,,
  \quad \Phi_{c}(\overline {M^2})\se 1.8\,.\eq
  Because we need in what follows the value of $f_D$ at the low normalization
  point $\mu_o$,\, we renormalize now both $r_D^2$ and $<{\bar q}\,q>$ to this
  point and (because they have the same anomalous dimension) obtain:
  \bq r_D^2(\mu_o^2)\se 2\,M_c\,<0|\,q\,{\bar q}\,|0>_{\mu_o}1.8\,\quad.\eq
  Finally,\, to obtain the concrete answer we use
  \bq <0|\,q\,{\bar q}\,|0>_{\mu_o}\se (\,0.25\,GeV\,)^3\,,\quad
  \mu_o^2\se 0.5\,GeV^2\,\,, \eq
  for the value of the quark condensate at the low normalization
  point. So,
  \bq r_D(\mu_o^2=0.5\,GeV^2)\se 0.30\,GeV^2\,,\quad f_D(\mu_o^2)
  \se 144\,MeV\,,& \nonumber \\
  & \displaystyle f_D(M_c^2)=\left (\frac{\alpha_s(\mu_o^2)}{\alpha_s
  (M_c^2)}\right )^{2/9}f_D(\mu_o^2)\se 165\,MeV\,.\eq

  For the B-meson,\, the corresponding sum rule has the form:
  \bq r_{B}^2(\overline {M^2})=2\,M_{b}\,<0|q\,{\bar q}|0>_{\overline {M^2}}\,
  \Phi_{b}(M^2)\,,\eq
  \bq \Phi_{b}(M^2)=exp \left \{ \frac{M_{B}^2-M_{b}^2}{M^2}\right \}\,
  \left [\,1-\kappa\,\frac{m_{o}^2\,M_{b}^2}{4\,M^4}\left (\,1-2\,\frac{M^2}
  {M_b^2}\right )\right ]\,. \eq
  The factor $\kappa$ in Eq.(125) $(\,\overline {M^2}\se 4\,GeV^2,\,\,
   \alpha_s(4\,GeV^2)\se 0.284\,)$ :
  \bq \kappa=\left (\frac{\alpha_s(\overline {M^2})}{\alpha_s(M_c^2)}\right )^
  {14/25}\left (\frac{\alpha_s(M_c^2)}{\alpha_s(1\,GeV^2)}\right )^{14/27}
  \se 0.81,\,\, \eq
  is due to different anomalous dimensions of the operators ${\bar q}
  ig_s\sigma G q$ and ${\bar q}q$\,.

  The corresponding interval (see above) of $M^2$ is: $3\,GeV^2\le
  M^2 \le 5\,GeV^2,\,$
  the function $\Phi_{b}(M^2)$ varies only slightly in this interval,\,
  and the characteristic value of $M^2$ (at the extremum) is:
  \bq \overline {M^2}\se 4\,GeV^2\,,
  \quad \Phi_{b}(\overline {M^2})\se 1.53\,.\eq

  Proceeding now in the same way as for the D-meson,\, we obtain:
  \bq r_B(\mu_o)\se 0.49\,GeV^2\,,\quad f_B(\mu_o)\se 89\,MeV\,, & \nonumber\\
  & \displaystyle f_B (M_b)=\left ({{\alpha_s(M_c^2)}
  \over{\alpha_s(M_b^2)}}\right )^{6/25}
  \left ({{\alpha_s(\mu_o^2)}\over{\alpha_s(M_c^2)}}\right )^
  {6/27}f_B(\mu_o)\se 113\,MeV\,. \eq

  It is of interest to compare the above results with those obtained
  in the static limit:\,$M_{q}\ra \infty.$ Let us define the constant\,
  $f_{o}$\, (a static analog of $f_{D}$) as:
  \bq f_{D}(\mu_o)= {{1}\over{\sqrt M_c}}\,f_{o}\,, \eq
  \bq M_D\se M_c+{\bar \Lambda}\,, \quad M^2\equiv 2\,M_c\,E\,,\quad
  {\bar \Lambda}\se 250\,MeV\,.\eq
  Then,\, in the limit $M_c\ra \infty$,\, the sum rule Eq.(118) takes the form
  \footnote{ Really,\, we drop out power corrections but keep the value of
  $m_o^2$ fixed.}:
  \bq f_{o}^2\se 2 <0|q\,{\bar q}|0>_{\mu_o} \phi (E)\,, \quad
  \phi (E)=e^{{\bar \Lambda}/E}\,\left (1-{{m_o^2}\over
  {16\,E^2}}\right )\,.\eq

  The corresponding interval (see above) of E is: \,\,$400\,MeV\le E \le
  700\,MeV,\,$ the function $\phi(E)$ varies only slightly in this region,\,
  and the characteristic value of E (at the extremum) is:
  \bq {\overline E}\se 550\,MeV\,, \quad \phi({\overline E}
  \se 550\,MeV)\se 1.32\,.\eq
  Substituting Eq.(132) into Eq.(131) we have:
  \bq f_{o}\se 0.20\,GeV^{3/2}\,,\quad f_{D}(M_c)\se \left [{{\alpha_{s}
  (\mu_o^2)}\over{\alpha_{s}(M_c^2}}\right ]^{2/9}\frac{1}{\sqrt {M_c}}
  \,f_o\se 180\,MeV\,. \eq
  Comparing Eq.(133) with Eq.(123) we see that the result obtained for
  $f_{D}$ in the static limit does not differ considerably from
  those obtained for the real value of the c-quark mass.

  Although we succeeded in using the correlator Eq.(115) in which both the
  Born contribution and all radiative corrections to it are absent,\, for the
  calculation of the decay constants $f_{D}$\, and $f_{B}$,\,
  the analogous trick,\, unfortunately,\, turns out to be useless for the
  calculation of nonfactorizable contributions to
  the matrix elements of the 4-quark operators which we
  will need in what follows. The reason is as follows. In the sum rules for
  the correlator Eq.(115) the leading contribution is proportional to the
  known quark condensate $<{\bar q}\,q>$,\, and the main correction is
  proportional to the condensate $<{\bar q}\sigma G q>$ which is also
  known. However,\, if we will try to calculate the matrix elements like
  (see below): $<D|{\bar c}\Gamma_{\mu}{\frac{\lambda^a}{2}}q\cdot {\bar q}
  \Gamma_{\nu}\frac{\lambda^a}{2}c|D>$ in an analogous way,\, then even the
  leading contribution is expressed through the high dimension condensate
  $<{\bar q}q{\bar q}qGG>$ which is unknown. Therefore,\, the only way to
  calculate such matrix elements with the help of the QCD sum rules is to use
  such interpolating currents that the leading contribution is pure
  perturbative,\, and the leading corrections are expressed then through the
  known low dimension condensates like $<G_{\mu\,\nu}^2>\,,$ etc.

As it was pointed out above,\, to deal with such sum rules we have to overcome
in some way the difficulties originating from very large radiative corrections.
We will describe now the way we used here to deal with this problem and which
is used then in sect.9.

  Let us return to a consideration of the decay constant $f_{D}$ and consider
  now the correlator:
  \bq K_{2}(\,q^2\,)=i \int d\,x\,e^{i\,q\,x}<O|T\left \{ {\bar Q}
  (x)(\,1+\gamma_{5}\,)\,q(x), \,{\bar q}(0)
  (\,1-\gamma_{5}\,)\,Q\,(0)\right \}|O>\equiv & \nonumber  \\
  & \displaystyle i\, \int d\,x\,e^{i\,q\,x}\,<O|T \left \{ P(x)\,P^{+}(0)+
  S(x)\,S^{+}(0)\,\right \}\,|O>\,,\eq
  The sum rule obtained from Eq.(134) in a standard way has the form (in the
  chiral limit):
  \bq r^{2}_{D}(\mu_o)\se \frac{3}{4\pi^2}\int^{S_{o}}_{M_{c}^{2}} {{{d\,
 S\,(\,S-M_{c}^{2}\,)^2}\over{S}}}\, exp \left \{\frac{M_D^2-S}{M^2}\right \}
  \left [ 1+\frac{4}{3}
  \,\frac{\alpha_{s}(\mu_o^2)}{\pi}\,X\left (\frac{M_c^2}{S}\right )
  +O(\alpha_s^2) \right ]\,,\eq
  \bq X(z)=\left [\, {{9}\over{4}}+2\,L(z)+\log (z)\,\log (1-z)+{{3}\over{2}}
  \,\log {{{z}\over{1-z}}}+\log {{{1}\over{1-z}}}- z\,\log {{{z}\over{1-z}}}+
  \right. & \nonumber \\
  & \displaystyle \left. {{z}\over{1-z}}\,
  \log {{{1}\over{z}}}-{{3}\over{4}}\,\log
  {{{M_c^2}\over{\mu_o^2}}} \right ] \,, \quad
  L(z)=- \int_{0}^{z}{{d\,t}\over{t}}\,\log (1-t)\,.\eq
  Let us emphasize that the chirality odd condensates like: $<{\bar q}q>\,,
  \, <{\bar q}\sigma G q>$ etc.,\, give no contribution
  to the correlator Eq.(134).
 \footnote{ We neglect the small contribution to Eq.(135) due to $<G^2>$ and
  $<{\bar q}\,q>^2$\,.}

  Let us try now,\, as a first approximation,\, to neglect all radiative
  corrections in Eq.(135) and let us choose the value of $S_o$ (the effective
  parameter which models the beginning of the perturbative continuum in a
  given correlator) in Eq.(135) to obtain a fit in $M^2$. There is a good fit
  in the standard region (see Eq.(120)\,): $0.8\,GeV^2\le M^2\le 1.5\,
  GeV^2\,$ at $S_o=3.81\,GeV^2$\,, which gives:
  \bq r_{D}(\mu_o)\se 0.096\,GeV^2\,,\quad f_{D}(\mu_o)\se 45\,MeV\,.\eq
  For the B-meson,\, there is a good fit in the standard region $3\,GeV^2\le
  M^2\le 5\,GeV^2$ at $S_o=28.8\,GeV^2$ which gives:
  \bq r_{B}(\mu_o)\se 0.19\,GeV^2\,, \quad f_{B}(\mu_o)\se 35\,MeV\,.\eq

  It is seen that the Born approximation to the sum rule Eq.(135) gives (at
  $M_c=1.65\,GeV,\,\,M_b=5.04\,GeV$) very small values for $f_{D},\, f_{B}$
  (compare with Eqs.(123),\,(128)).
  Let us account now for the one loop correction
  to the Born approximation in the sum rule Eq.(135) and make anew the fits in
  $M^2$. We will obtain good fits both for D- and B-mesons in the standard
  regions of $M^2$-values at $S_o=3.85\,GeV^2$ and $S_o=28.9\,GeV^2$
  respectively with the results:
  \bq r_{D}(\mu_o)\se 0.149\,GeV^2\,,\quad f_{D}(\mu_o)\se 71\,MeV\,.\eq
  \bq r_{B}(\mu_o)\se 0.324\,GeV^2\,, \quad f_{B}(\mu_o)\se 59\,MeV\,.\eq
  It is seen from a comparison of
  Eqs.(139),(140) and Eqs.(137),(138) that if,\, in the
  Born approximation,\, we obtain the values of $f_{D},\,f_{B}$ which are
  $\se 3$ times smaller than the right values Eqs.(123),\,(128),\,
  on account of
  the first radiative correction the results increase $\se 1.6-1.7$ times.
  It is clear that,\, in such a situation,\, one needs either to account for
  all radiative corrections,\, or to use some trick in the hope that it can
  help to account effectively for a summary effect of all radiative
  corrections with reasonable accuracy. We have chosen the second way,\,
  of course.

  The large radiative corrections in the correlator Eq.(134) look reasonable
 as they tend to increase the Born contribution which is too small. On the
  other hand,\, these large corrections indicate that the values of the
  parameters entering the spectral density in Eq.(134) are not chosen
  properly. In the given case,\, the only parameter is the quark mass,\,
  $M_c$,\, (we used everywhere above the "hard pole masses"),\, and the
  correlator Eq.(134) is very sensitive to the
  precise value of the quark mass (with the D-meson mass fixed). The right
  hand side of Eq.(135) increases greatly when the quark mass decreases.

  Therefore,\, it looks natural to try to describe the main effect of radiative
  corrections by using the effective quark masses: $\mu_c\le M_c,\, \mu_b\le
  M_b.$ So,\, let us define:
  \bq M_c=\mu_c\left (1+{{\alpha_s(\mu_o^2)}\over{4\,\pi}}\,
  C_D+O(\alpha_{s}^2) \right ) \eq
  (and analogously for $M_b$),\, and let us express $M_c$ through $\mu_c$ at
  the right hand side of Eq.(135). We obtain:
  \bq r^{2}_{D}(\mu_o)\se \frac{3}{4\pi^2}\int^{S_{o}}_{\mu_{c}^{2}} \frac
 {d\,S\,(\,S-\mu_c^2\,)^2}{S}\, exp \left \{\frac{M_D^2-S}{M^2}\right \}
  \left [ 1+{{4}\over{3}}\,{{\alpha_{s}(\mu_o^2)}\over{\pi}}\,{\bar X}
  ({{\mu_c^2}\over{S}})+O(\alpha_{s}^2) \right ]\,,\eq
  \bq {\bar X}(z)=X(z)-{{3}\over{4}}\,C_D\,{{z}\over{1-z}}\,.\eq
  Let us try now to find a value of $\mu_c$ which will give the answer for
$f_D$ close to the right one,\, Eq.(123),\,  and for which the radiative
corrections will remain reasonably small at the same time. It is clear that
there is no guaranty that it is possible to succeed in this way.

  Nevertheless,\, the results are very encouraging. For instance,\, let us
  suppose that the radiative corrections are reasonably small (for a
  properly chosen value of $\mu_c$),\, put ${\bar X}(z)=0$ in Eq.(142) and find
  the value of $\mu_c$ which reproduces the right answer,\, Eq.(123). For:
  $\mu_c=1.40\,GeV,\,\,S_o=4.4\,GeV^2$ one obtains a good fit in the
  standard region $0.8\,GeV^2\le M^2\le 1.5\,GeV^2$ with the result $r_
  D(\mu_o)\se 0.3\,GeV^2$ (compare with Eq.(123)). To elucidate the role of
  radiative corrections in the sum rule Eq.(142),\, let
  us calculate it now in the same region and with the same papameters:
  $\mu_c=1.40\,GeV,\,S_o=4.4\,GeV^2$,\, and $C_D=3.87$\,
  (see Eq.(141)). The fit is not optimal but sufficiently good,
  and when the right hand side of Eq.(142)
  is taken at the characteristic value $\overline {M^2}=1.15\,GeV^2$ it gives:
  $r_D(\mu_o)\se 0.29\,GeV^2,$\, in good agreement with Eq.(123).

  For the B-meson,\, the situation is analogous but slightly worse.
  Namely,\, let us choose $\mu_b=4.84\,
  GeV$ and put ${\bar X}(z)=0$. We obtain then a good fit in the standard
  region: $3\,GeV^2\le M^2\le 5\,GeV^2$ at $S_o=30.0\,GeV^2$ with the
  result: $f_{B}(\mu_o)\se 89\,MeV$,\, which
  reproduces Eq.(128). Accounting now for the radiative correction,\, using
  $\mu_b=4.84\,GeV,\, S_o=30\,
 GeV^2,\,$ and $C_B=0.895$\, (see Eq.(141)),\, and taking the right hand side
 at the characteristic value $\overline {M^2}=4\,GeV^2$,\, we obtain the result
 for $r_B(\mu_o)$ which is $\sim 15\%$ higher than the right value,\, Eq.(128).
 Fortunately,\, because the power corrections to the B-meson decay widths
  are small (see below),\, such accuracy will be sufficient for
  our purposes.

  In summary,\, it is possible to reproduce the right values of $f_D$ and
  $f_B$ by neglecting the radiative corrections in the sum rule Eq.(142) and
  using the effective quark masses:
  \bq \mu_{c}=1.40\,GeV\,,\quad \mu_{b}=4.84\,GeV\,.\eq
  The residual effect of radiative corrections remains reasonably small in
  this case.

  As for the correlator Eq.(134) and the decay constants $f_D$ and $f_B$,\,
  there was no need to perform all the above manipulations because the answer
  was obtained previously in this section. As it was pointed out above,\,
  our real
  purpose is to calculate more complicated matrix elements of the 4-quark
  operators (see sect.9). And our main assumption is that for the correlators
  used below in sect.9 it is possible to obtain reasonable results by
  neglecting radiative corrections and using $\mu_c$ and $\mu_b$ instead of
  $M_c$ and $M_b$ in the corresponding spectral densities.\\

{\bf 8\,.  Difficulties with naive estimates.} \\

 Let us turn now to $\Gamma^{o}_{nl}$ defined in Eq.(25).
 All the quantities entering are known (because the radiative corrections
 are known here for $m_s=0$ only and we neglect below the SU(3)-symmetry
breaking corrections,\, we put $m_s=0$ in nonleptonic calculations):
 \bq \left ({{2C_{+}^2+C_{-}^2}\over{3}}\right )\se 1.43\,,\quad \left (1-2\,
 \Delta_{G}\right )\se 0.7\,, \quad 4\left ({{C_{-}^2-C_{+}^2}\over
 {2C_{+}^2+C_{-}^2}}\right )\Delta_{G}\se 0.36\,,\eq
 \bq {{<D|{\bar c}c|D>}\over{2\,M_D}}\se 1.02\,, \quad
 I_{rad}\se 1.05\,,\quad \Gamma_{o}={{G_F^2\,
 M_c^5}\over{64\,\pi^3}}\se 8.4\cdot 10^{-13}\,GeV\,,\eq
 \bq \Gamma_{nl}^o \se \Gamma_{o}\bigl [1.43 \bigr ]_{rad}\,
\bigl [1.02 \bigr ]_{{\bar c}c}\Biggl \{\,\bigl [1.05 \bigr ]_{rad}
\biggl (\,1-\bigl [0.3\bigr ]_{\sigma G}\biggr )+\bigl [0.36 \bigr ]_
{\sigma G}\,\Biggr \}\se 1.59\,\Gamma_o\,.\eq
 Let us try now to obtain a rough estimate of the matrix element\,
 $<D^{+}|\delta L_{eff}|D^{+}>,\,$ see Eq.(45),\, by putting:
 $\lambda\se P_D$ in $L_s,\,L_u$ (see fig.3,\, $P_D$ is the D-meson
momentum),\,
 ${\bar \lambda}^2\se P_c^2\se M_c^2$ in $L_d$ and using the
 factorization approximation \cite{VS},\,\cite{BS2}.
 Then,\, $L_u$ and $L_s$ give zero contributions and:
 \bq  <D^+(p)|\,{\bar c}\,\Gamma_{\mu}\,d \cdot {\bar d}\,\Gamma_{\nu}\,c\,|
 D^+(p)>_{\mu_o}\se p_{\mu}\,p_{\nu}\,f_D^2(\mu_o)\,,\eq
 \bq  <D^{+}|\,L_d\,|D^{+}>_{\mu_o}^{factor}
 \se -1.1\,M_c^2\,f_D^2(\mu_o)\,M_D^2\,,\eq
 \bq \Delta \Gamma_{factor}(D^{+})\se \left [ -1.1\cdot 16\,\pi^2\,
 {{f_D^2(\mu_o)\,M_D}\over{M_c^3}}\right ]\,\Gamma_o\se -1.50\, \Gamma_o\,.\eq
 The contribution from $L_{PNV}$ in Eq.(51) is obtained by using:
 \bq <D|\,J^a_\rho\cdot {\bar c}\,\gamma_\rho (1\pm \gamma_5){{\lambda^a}
 \over{2}}\,c\,|D>^{factor}_{\mu_o}\se -{{4}\over{9}}\,r_D^2(\mu_o)
 \left (\,1-{{M_c^2}\over{2\,M_D^2}}\,\right )\,,\eq
 \bq {{\Delta \Gamma_{PNV}}\over{\Gamma_o}}
 \se -{{64}\over{9}}\pi^2\,\,\frac{f_D^2(\mu_o)\,M_D^3}{M_c^5}\,N_v
 \left (1-\eta_{co}^{-2/9}\right )\left (\,1-{{M_c^2}\over
 {2\,M_D^2}}\,\right )\se -0.15\,.\eq
 On the whole, one obtains:
 \bq  \Gamma_{nl}(D^+)\se \left (1.59-1.50-0.15 \right )\,\Gamma_o \se
 -0.06\,\Gamma_o\,,\eq
 which does not make much sense. It is clear that the above
 approximations are too rough and some estimates are essentially wrong. It
 is the purpose of subsequent sections to improve the above described
 naive estimates.\\

   {\bf 9\,.  Nonfactorizable contributions: gluon condensates}. \\

 We calculate in this section the nonfactorizable gluon
 contributions to the matrix elements of the 4-quark operators. Let
 us begin with the operator $O_{\mu\,\nu}={\bar c}\Gamma_{\mu}(\lambda^a/2)q
 \cdot {\bar q}\Gamma_{\nu}(\lambda^a/2)c$,\, which gives zero matrix elements
 in the factorization approximation. We want to calculate the contribution
 of the fig.6 diagrams with the help of the QCD sum rules.

 \begin{center} \{Fig.\,6a\} \end{center}

 \begin{center} {\bf Fig.\,6a}\,\,\,\, The nonfactorizable gluon contribution
 to the weak annihilation \end{center}

 \begin{center} \{Fig.\,6b\} \end{center}

 \begin{center} {\bf Fig.\,6b}\,\,\,\, The nonfactorizable gluon contribution
 to the cross weak annihilation \end{center}

 \begin{center} \{Fig.\,7\} \end{center}

 \begin{center} {\bf Figs.\,7a,\,7b}\,\,\,\, The diagrams for the meson
 transition into a current with an emission of a nonperturbative gluon
 \end{center}

 Practically,\, however,\, it is more convenient to calculate the
 nonperturbative gluon emission amplitude,\,
 fig.7,\, and to obtain then the contribution to the decay width by averaging
 its modulus squared over the gluon field fluctuations in the QCD vacuum.

 To calculate the fig.7 amplitude,\, we replace the D-meson by the
 interpolating current and consider the correlator:
  \bq T^{\mu}_{1}=i\,\int d\,x\,exp\,\{ipx \}\,<1\,gl\,|T\,J^a_{\mu}(x)\,J_{P}
 (0)\,|0>_{\mu_o}\,, \eq
 \bq J^a_{\mu}(x)={\bar q}\,(x)\,\Gamma_{\mu}\,{{\lambda^a}\over{2}}\,
 c(x)\,, \quad J_{P}(0)={\bar c}\,(0)i(\,1+\gamma_5\,)\,
 q(0)\,.\eq
 Calculating the contributions of figs.7\,a,\,b diagrams to the
 discontinuity of $T^{\mu}_1$ in $p^2,\,$ integrating
 it from the threshold up to $S_o$ with the weight\,\,
 $exp \{-p^2/M^2 \}$\, and equating to the D-meson contribution,\, we
 obtain the sum rule for the D-meson transition amplitude into the
 current $J^a_{\mu}$ with an emission of a nonperturbative gluon
 \footnote {The term $I_G$ in Eq.(156) is due to the fig.7b diagram and is
 parametrically smaller than those from fig.7a.}:
 \bq N_{\mu}\se {{4\pi M_{c}\,g_{s}}\over{(\,2\,\pi\,)^4\,r_{D}(\mu_{o})}}
 \left \{i \,{\widetilde G}^a_{\mu\,\alpha}\,{\widetilde I}_{G}+\,G^a_
 {\mu\,\alpha}\,I_{G}\right \}\,p_{\alpha}\,, \eq
 \bq {\widetilde I}_{G}=\int^{S_{o}}_{\mu_{c}^{2}} {{d\,S}\over{S}}\,exp \left
 \{{{M_D^2-S}\over{M^2}}\right \}\,;\quad
 I_{G}=\int^{S_{o}}_{\mu_{c}^{2}} {{d\,S\,(\,S-\mu_{c}^{2}\,)}\over
 {S^2}}\,exp {\left \{{{M_D^2-S}\over{M^2}}\right \}}\,. \eq
 \bq {\widetilde G}_{\mu\,\nu}={{1}\over{2}}\,\epsilon_
  {\mu\,\nu\,\lambda\,\sigma}\,G_{\lambda\,\sigma}\,.\nonumber \eq
 (In accordance with the discussion in sect.7,\, we replaced $M_c$ by
 $\mu_c$ in the spectral density,\, see Eq.(144),\, supposing this accounts for
 the major effect of radiative corrections).

 Unfortunately,\, the attempt fails to obtain a good fit for ${\widetilde I}
^c_{G}$ in the standard region of $M^2$: $0.8\,GeV^2\le M^2\le 1.5\,GeV^2$,\,
 and with $S_o$ in the reasonable vicinity of its optimal value
 $S_o=4.4\,GeV^2$ (see sect.7),\, because ${\widetilde I}^c_G$
 is nearly independent of $S_o$. For
 instanse,\, ${\widetilde I}^{c}_{G}$ varies as: $2.0\ge {\widetilde I}
 ^{c}_{G}\ge 1.2$ in this region at $S_o=4.4\,GeV^2$. Therefore,\, the best
 we can do is to take for ${\widetilde I}^{c}_{G}$ its value at the
 characteristic point $\overline {M^2}=1.15\,GeV^2$\, (see sect.7)\, :
 \bq {\widetilde I}^{c}_{G}\se {\widetilde I}^{c}_
 {G}(\overline {M^2}=1.15\,GeV^2)\se 1.4\,.\eq
 Analogously,\, $0.4\ge I^{c}_{G}\ge 0.3$ in the standard interval of
 $M^2$ and at $S_o=4.4\,GeV^2\,.$ So,\, we take:
 \bq I^{c}_{G} \se I^{c}_{G}\,(\overline {M^2}=1.15\,GeV^2)\se 0.33\,.\eq
 On the whole,\, we estimate:
 \bq N_{\mu}\se {{4\pi M_{c}\,g_{s}}\over{(\,2\,\pi\,)^4\,r_{D}(\mu_{o})}}
 \left \{\,1.4i\,{\widetilde G}^a_{\mu\,\alpha}+\,0.33\,G^a_{\mu\,\alpha}
 \,\right \}\,p_{\alpha}\,. \eq
 Taking now the product $N_{\mu}\,N^{+}_{\nu}$ and averaging the gluon fields
 over the vacuum,\, we have finally \footnote{ It is implied,\, see figs.6,7,
\, that the quark flavour in Eq.(161) is such that it is a valent one.}:
 \bq <D(p)|\left [{\bar c}\Gamma_{\mu}{{\lambda^a}\over{2}}\,q \cdot {\bar q}
\Gamma_{\nu}{{\lambda^a}\over{2}}\,c\right ]_{\mu_o}|D(p)>_{gl}\,\se \left
({{p_{\mu}\,p_{\nu}}\over{p^2}}-g_{\mu\,\nu}\right )\,C^{c}_{G}\,,\eq
 \bq C^{c}_{G}\se {{M_{D}^{2}\,M_{c}^{2}}\over{48\,\pi^2\,r_{D}^{2}(\mu_o)}}\,
 <0|{{\alpha_{s}}\over{\pi}}\,G_{\mu\,\nu}^{2}|0>\left
 [(\,{\widetilde I}^c_G\,)^2-(\,I^c_G\,)^2
 \right ] \se 0.48\cdot 10^{-2}\,GeV^4\,.\eq
 The situation is qualitatively the same for the B-meson (${\widetilde I}^b_G$
 varies as: $0.46\ge {\widetilde I}^b_G\ge 0.35$ in the standard region
 $3\,GeV^2\leq M^2\leq 5\,GeV^2,\,\, S_o=30\,GeV^2$),\,
 and proceeding in the same way we obtain
${\widetilde I}^b_G\se 0.4,\,\,\, I^b_G\se 0.03\,,$\, so that:
 \bq C^{b}_G\se 1.15\cdot 10^{-2}\,GeV^4\,.\eq

 Let us point out that the parametric behaviour of ${\widetilde I}_G$ is:
 ${\widetilde I}_G=0(\mu_o/M_c)\,,\,I_G=0(\mu_o^2/M_c^2).$ So,\, the
matrix element Eq.(161) which describes the non-perturbative contribution to
WA (weak annihilation,\, fig.6a ) and CWA (cross weak annihiltion,\, fig.6b )
behaves as: $0({\mu_o^3}\,M_c),\,$ i.e. in the same way as the factorizable
contribution Eq.(148). This gives a relative correction to the decay width:
${\delta \Gamma}/{\Gamma}=0(\mu_o^3/M_c^3)$,\, as it should be.

In order to understand to what extent the factorization approximation is
 good,\, let us compare Eqs.(161)-(163) with the factorizable matrix elements:
 \bq <D(p)|{\bar c}\Gamma_{\mu}\,q \cdot {\bar q}\Gamma_{\nu}\,c
 |D(p)>^{fact}_{\mu_o}\se \left ({{p_{\mu}\,
 p_{\nu}}\over{p^2}}\right )
 f_{D}^{2}(\mu_o)M_{D}^{2}\se \left ({{p_{\mu}\,p_{\nu}}\over{p^2}}\right )
 7.2\cdot 10^{-2}\,GeV^4\,.  \eq
\bq <B(p)|{\bar b}\Gamma_{\mu}\,q \cdot {\bar q}\Gamma_{\nu}\,b
 |B(p)>^{fact}_{\mu_o}\se \left ({{p_{\mu}\,
 p_{\nu}}\over{p^2}}\right )
 f_{B}^{2}(\mu_o)M_{B}^{2}\se \left ({{p_{\mu}\,p_{\nu}}\over{p^2}}\right )
 22.0\cdot 10^{-2}\,GeV^4\,. \nonumber \eq
It is seen that the nonfactorizable contributions are $\se 15-20$ times smaller
than the corresponding factorizable one,\, so that the factorization
approximation works very well even for the D-mesons.

It seems therefore that there are no chances to change essentially the
results obtained in the previous section. As will be shown below,\, this
is not the case really for the following reasons:\\
1) Although the matrix element Eq.(161) is small,\, it enters $\delta L_{eff}$
with much larger coefficients in comparison with the
factorizable contributions;\\
2) As will be shown in sect.12,\, the characteristic value of ${\bar
\lambda^2}$ in $L_d$\,, Eq.(31) is: $<{\bar \lambda}^2>\se 0.35\,M_D^2,\,$
while
those of ${\lambda^2}$ in $L_u,\,L_s$ is: $<\lambda^2>\se M_D^2,\,$ and this
effect suppresses strongly the large factorizable contribution,\, Eq.(164).

Let us perform now an estimate of the nonfactorizable contribution to the
matrix element of the operator
 $S_{\mu\,\nu}={\bar c}\Gamma_{\mu}\,q \cdot {\bar q}\Gamma_{\nu}\,c$\,.
For this,\, let us consider the correlator:
 \bq T^{\mu}_{2}=i\,\int d\,x\,exp\,\{ipx \}\,<2\,gl\,|T\,J_{\mu}(x)\,J_{P}
 (0)\,|0>_{\mu_o}\,, \eq
 \bq J_{\mu}(x)={\bar q}\,(x)\,\Gamma_{\mu}\,
 c(x)\,, \quad J_{P}(0)={\bar c}\,(0)i(\,1+\gamma_5\,)\,q(0)\, \eq
 and calculate the fig.8a contribution.
 \footnote {Because we can expect
 beforehand (see the above discussion) that the nonfactorizable correction
 will be small,\, we confine ourselves to the main contribution from the
 fig.8a diagram,\, neglecting even smaller contributions from those
 diagrams where gluons are emitted by the c-quarks.} We have:
 \bq T^{\mu}_2={{8i\,M_{c}\,g_{s}^{2}}\over{(\,2\,\pi\,)^{4}}}\int
 {{d\,k}\over{k^{8}}\,[(\,p+k\,)^2-M_c^2]}\left (k^{2}\,k_{\nu}\,g_{\mu\,\rho}-
 k_{\mu}\,k_{\nu}\,k_{\rho}\,\right )\,\left [\,G_{\rho\,\lambda}^{a}\,
 G_{\lambda\,\nu}^{a}\,\right ]\,.\eq
 One obtains from Eq.(167) with logarithmic accuracy:
 \bq T_{2}^{\mu}\se \Omega_{\mu}\,{{2}\over{3}}\,M_{c}\,\log \left ({
 {\mu_{max}^{2}}\over{\mu_{min}^2}}\right )\,{{\partial}\over{\partial p^2}}\,
 {{1}\over{p^2-M_{c}^{2}}}\,,\eq
 \bq \Omega_{\mu}={{\alpha_{s}}\over{\pi}}\,\left [\,G_{\mu\,\lambda}^{a}\,
 G_{\rho\,\lambda}^{a}\,p_{\rho}-{{1}\over{4}}\,p_{\mu}\,G_{\rho\,\lambda}^{a}
 \,G_{\rho\,\lambda}^{a}\,\right ]\,.\eq
 Proceeding now in the standard way,\, one obtains the sum rule for the
 D-meson transition amplitude into the current $J_{\mu}$ and two gluons:
 \bq M_{\mu}\se \Omega_{\mu}\,\frac{2}{3} M_c\,\frac{1}{r_D(\mu_o)}\,
\log \left (\frac{\mu_{max}^2}{\mu_{min}^2}\right )\,\frac{1}{M^2}\,exp
\left \{\frac{M_D^2-M_c^2}{M^2}\right \}+ \cdots \eq
 As we need here a rough estimate only,\, let us put:
 $\overline {M^2}=1\,GeV^2$ in Eq.(170) (see sect.7).
 Besides,\, because both $\mu^2_{max}$ and $\mu^2_{min}$ in Eq.(170)
 remain finite at $M_c\ra \infty$,\, we put:
 \bq \log \left ({{\mu_{max}^{2}}\over{\mu_{min}^2}}\right )\se 1. \eq
 Therefore,
 \bq M_{\mu}\sim \left (\,7.9\,GeV^{-3}\right )\,\Omega_{\mu}\,.\eq
 Now,\, taking the product $M_{\mu}\,M^{+}_{\nu}$ and averaging the gluon
fields
 ,\, one has:
 \bq <D(p)|{\bar c}\Gamma_{\mu}\,q
 \cdot {\bar q}\Gamma_{\nu}\,c\,|D(p)>_{nonfact}
 \se \left (\,62\,GeV^{-6}\,\right )\,<0|\Omega_{\mu}\,\Omega_{\nu}\,|0>.\eq
 Unfortunately,\, this 4-gluon vacuum condensate is unknown. So,\, let us
obtain
 first an estimate in the factorization approximation,\, see fig.8b.

 \begin{center} \{Figs.\,8a,\,8b\} \end{center}

 \begin{center} {\bf Fig.\,8}\,\,\,\, The nonfactorizable contribution to
 the matrix element $<D|S_{\mu\,\nu}|D>$ \end{center}

One has:
 \bq <0|\,\Omega_{\mu}\,\Omega_{\nu}\,|0>
 \sim {{1}\over{576}}\,g_{\mu\,\nu}M^2_D\,
 <0|{{\alpha_s}\over{\pi}}\,G_{\mu\,\nu}^2|0>^{2}+O(p_{\mu}\,p_{\nu})\,,\eq
 so that:
 \bq <D(p)|\,{\bar c}\,\Gamma_{\mu}\,q
 \cdot {\bar q}\,\Gamma_{\nu}\,c\,|D(p)>_{nonfact}\sim g_{\mu\,\nu}\left (
 0.5\cdot 10^{-4}\,GeV^4\right )\,,\eq
 (and the term $\sim p_{\mu}\,p_{\nu}$ is of the same order). Comparing
 Eq.(175) with Eqs.(164) and (161) we see that the nonfactorizable
 correction is very small in
 this matrix element. Therefore,\, even if the approximations made above give
 the right order of magnitude only,\, we can safely neglect this correction.\\

{\bf 10\,.  Nonfactorizable contributions: quark condensates}. \\

The contribution $<D|\delta L_{PNV}|D>$ was calculated in sect.8 in the
factorization approximation. Because it is not large by itself and
nonfactorizable corrections are also small (see sect.9),\, it does not make
much sense to account for them in this matrix element. So,\, we will use
for it the expression Eq.(152).

There are,\, however,\, analogous corrections of the "penguin" type in the
matrix elements of the operators $L_u,\, L_d,\, L_s$ in Eqs.(45)-(48),\,
see figs.9,\,10,\, and we proceed now to their calculation.

\begin{center} \{Fig.\,9\} \end{center}

\begin{center} {\bf Fig.\,9}\,\,\,\, The nonvalence nonfactorizable penguin
 contribution (NV) \end{center}

 \begin{center} \{Fig.\,10\} \end{center}

 \begin{center} {\bf Fig.\,10}\,\,\,\, The valence nonfactorizable penguin
 contribution \end{center}

 For a calculation of the diagrams
 in figs.9,\,10 (see also fig.4) let us consider the correlator:
\bq T_{\mu\,\nu}=i\int d\,x\,e^{ip_1x}\,i\int d\,y\,e^{-ip_2y}
<0|T\,J_P^{+}(x)\,S_{\mu\,\nu}(0)\,J_P(y)|0>,\,\eq
\bq J_P={\bar c}\,i\,(1+\gamma_5)q,\quad J^{+}_P={\bar
q}\,i\,(-1+\gamma_5)c,\quad
S_{\mu\,\nu}={\bar c}\Gamma_{\mu}\psi\cdot {\bar \psi}\Gamma_{\nu}c\,,\eq
where q and $\psi$ are light quark fields.

One can neglect the co-ordinate dependence of the light quark fields in
Eq.(176) for calculations with logarithmic accuracy,\, and obtain:
\bq T_{\mu\,\nu}\se \left ({{2\,M_c}\over{p_1^2-M_c^2}}\right )\,
\left ({{2\,M_c}\over{p_2^2-M_c^2}}\right ) K_{\mu\,\nu}+\cdots\,,\eq
\bq K_{\mu\,\nu}=<0|{\bar q}(0)\Gamma_{\mu}\psi(0)\cdot {\bar \psi}(0)\Gamma_{
\nu}q(0)|0>.\eq
Proceeding in the usual way one obtains now from Eq.(178) the sum rule:
\bq <D(p)|S_{\mu\,\nu}(0)|D(p)>_{peng}\se K_{\mu\,\nu} \left [
{{2\,M_c}\over{r_D}}\, e^{{{M_D^2-M_c^2}\over{M_1^2}}}+...\right ]\left [
\frac{2\,M_c}{r_D}\, e^{\frac{M_D^2-M_c^2}{M_2^2}}+...\right ]\,,\eq
where the dots denote power corrections in $1/M_1^2,\,1/M_2^2$.

Let us consider now the contribution of various penguin-diagrams to the
matrix element $<D|\delta L_{eff}|D>,\,$ (see Eq.(45)).
In order to check the normalization,\, let us start with the contribution of
fig.4\,. Separating in $K_{\mu\,\nu},\,$ Eq.(179),\,
the penguin contribution from the operator $\psi\,{\bar \psi}$:
\bq  \left [ \psi\,{\bar \psi}\right ]_{peng}\ra
\frac{1-\eta_{co}^{-2/b_o}}{6}
\left (\frac{\lambda^a}{2}\,\gamma_{\rho}\right )\,J^{a}_{\rho}(3)\,, \eq
and factorizing in the standard way the 4-quark condensate,\, we obtain:
\bq <D|S_{\mu\,\nu}|D>_{fig.4}\se \frac{2}{27}\,g_{\mu\,\nu}
\,(1-\eta_{co}^{-2/b_o}) & \nonumber \\ & \displaystyle \left [
\frac{2\,M_c <q{\bar q}>}{r_D}\, e^{\frac{M_D^2-M_c^2}{M_1^2}}+...
\right ]\left [\frac{2\,M_c <q{\bar q}>}{r_D}\, e^{\frac{M_D^2-M_c^2}
{M_2^2}}+...\right ].\eq
On the other hand,\, to calculate this contribution there is no need to use the
sum rules at all because,\, after the replacement Eq.(181),\, this contribution
to the matrix element can be obtained directly through factorization:
\bq <D|\,S_{\mu\,\nu}\ra
\left ( {{1-\eta_{co}^{-2/b_o}}\over{6}}\right )\,
{\bar c}\Gamma_{\mu}{{\lambda^a}\over{2}}\gamma_{\rho}\Gamma_{\nu}c\cdot J^a_
{\rho}\,|D>_{fig.4}\se {{2}\over{27}}\,g_{\mu\,\nu}\left
 (1-\eta^{-2/b_o}\right ) r_D^2\,.\eq
Comparing with Eq.(182) we see that we simply reproduced the
sum rule used before (see Eqs.(117),(118)):
\bq 2\,M_c\,<0|\,q\,{\bar q}\,|0>\,e^{{{M_D^2-M_c^2}\over{M^2}}}\left [1+\cdots
\right ]\se r_D^2\,.\eq
Therefore,\, we can rewrite Eq.(180) in the form:
\bq <D|\,S_{\mu\,\nu}\,|D>\se {{r_D^2}\over{<0|{\bar q}q|0>^2}}\,
K_{\mu\,\nu}\,.\eq
It is not difficult to obtain now other penguin contributions from Eq.(185).
Namely,\, the contribution of fig.9 is obtained by separating out the penguin
contribution from the operator $[q\,{\bar q}]$ in $K_{\mu\,\nu}$:
\bq [\,q\,{\bar q}\,]_{peng}\ra {{1}\over{6}}\left [{{\alpha_s(\mu_o^2)}\over
{2\,\pi}}\,\log \left ({{\mu_{max}^2}\over{\mu_{min}^2}}
\right )\right ]\,\left ({{\lambda^a}\over{2}}\right )\, J_{\rho}^a\,,\eq
which gives after the standard 4-quark condensate factorization:
\bq <D|\left [ S_{\mu\,\nu}\right ]_{\mu_o}\,|D>_{fig.9}\se
{{2}\over{27}}\,g_{\mu\,\nu}\,r_D^2(\mu_o)\left [{{\alpha_s
(\mu_o^2)}\over{2\,\pi}}\,\log \left ({{\mu_{max}^2}\over
{\mu_{min}^2}}\right )\right ]\Delta_s\,,\eq
where $\Delta_s=1$ if the $\psi$ in $S_{\mu\,\nu}$ is u- or d-quark,\,
and $\Delta_s=(<{\bar s}s>/<{\bar u}u>)^2\se 0.64$ if $\psi$ is the s-quark.
Let us point out that,\, unlike the standard penguin contribution,\, fig.4,\,
which contains the factor: $(1-\eta_{co}^{-2/b_o})\se \left (\alpha_s(\mu_o^2)
/2\pi)\log(M_c^2/\mu_o^2)\right )$,\, the fig.9 contribution
contains as an upper cut off the quantity $\mu_{max}^2$ which
remais finite at $M_c\ra \infty$ (see also Eq.(171)).\footnote
{It is clear that the contributions of figs.4,\,9 are nonvalence,\, i.e.
the same for all $D^{o,\pm,s}$-mesons;\, they only
shift the position of the "decay width centre"\,\,$\Gamma^{o}_{nl}$.}

If we consider now the operator $O_{\mu\,\nu}=[\,
{\bar c}\frac{\lambda^a}{2}\Gamma_{\mu}\psi \cdot {\bar \psi}
\frac{\lambda^a}{2}\Gamma_{\nu}c\,]$ instead of $S_{\mu\,\nu}$,\,
Eq.(177),\, then we have clearly:
\bq <D|O_{\mu\,\nu}|D>_{fig.9}=
\,-\,{{1}\over{6}}<D|S_{\mu\,\nu}|D>_{fig.9}\,.\eq
For the operator $O_{\mu\,\nu},\,$ however,\, there is an additional (valence)
penguin contribution originating from the diagram fig.10 (plus the mirror
one). It can be obtained easily from the relation analogous to Eq.(185):
\bq <D|\,O_{\mu\,\nu}\,|D>\se {{r_D^2}\over{<0|{\bar q}q|0>^2}}\,
P_{\mu\,\nu}\,.\eq
\bq P_{\mu\,\nu}=<0|\,{\bar q}\,\Gamma_{\mu}{{\lambda^a}\over{2}}q \cdot
{\bar q}\,\Gamma_{\nu}{{\lambda^a}\over{2}}q\,|0>\,, \eq
by separating out the penguin contribution:
\bq \left [ {\bar q}\Gamma_{\mu}{{\lambda^a}\over{2}}q\right ]_{peng}\ra
\left [-\,{{\alpha_s(\mu_o^2)}\over{6\,\pi}}\log \left ({{\mu_{max}^2}\over
{\mu_{min}^2}}\right)\right ] J^{a}_{\mu}\,,\eq
and factorizing the 4-quark condensate. We obtain (the factor 2
accounts for the mirror diagram):
\bq <D|\left [ O_{\mu\,\nu}\right ]_{\mu_o}\,|D>_{fig.10}\se {{2}\over{27}}\,
g_{\mu\,\nu}\,r_D^2(\mu_o)\left [{{\alpha_s(\mu_o^2)}\over{2\,\pi}}\,\log
\left ({{\mu_{max}^2}\over{\mu_{min}^2}}\right )\right ]\,.\eq
Let us point out finally that,\, within the logarithmic approximation,\,
we can put: $\lambda\se {\bar \lambda}\se P_c\se M_c$ for the contribution
of fig.4,\, and $\lambda\se {\bar \lambda}\se P_D$ for those of figs.9,\,10\,.

For the B-mesons,\, it is sufficient to replace: $r_D(\mu_o)\ra r_B(\mu_o)$ in
the above formulae Eqs.(187),\,(188),\,(192) to obtain contributions to the
matrix elements of ${\bar L}_u,\,L_d,\,L_s$ in Eq.(61) (for the $B^{\pm,o,s}$
mesons there are no contributions from $L_c,\,{\bar L}_c,\,{\tilde L}_c$
in Eq.(61)). \\

{\bf 11.\,\quad   Corrections to semileptonic widths\,.} \\

We are ready now to explain how the value of $\delta^{lept}(D^+)$ used in
sect.3 was obtained.

Let us recall (see sects.8,\,10 and fig.9) that:
\bq  <D^{+}(p)|\,{\bar c}\,\Gamma_{\mu}\,s\cdot
{\bar s}\,\Gamma_{\nu}\,c\,|D^{+}(p)>_{\mu_o}\se \frac{2}{27}g_{\mu\,\nu}
\,\rho_o\,\left ( \frac{<{\bar s}s>}{<{\bar u}u>}\right )^2
\,r_D^2(\mu_o)\,, & \nonumber \\
& \displaystyle <D^{+}(p)|\,{\bar c}\frac{\lambda^a}{2}\Gamma_{\mu}\,s\cdot
{\bar s}\frac{\lambda^a}{2}\Gamma_{\nu}\,c\,|D^{+}(p)>_{\mu_o}\se -\frac{1}{6}
<D^{+}(p)|\,{\bar c}\,\Gamma_{\mu}\,s\cdot
{\bar s}\,\Gamma_{\nu}\,c\,|D^{+}(p)>_{\mu_o}\,, & \nonumber \\
 & \displaystyle \rho_o=\frac{\alpha_s(\mu_o^2)}{2\,\pi}\,\log \left (
 \frac{\mu_{max}^2}{\mu_{min}^2}\right )\se 0.1\,, \quad
\frac{<0|{\bar s}s|0>}{<0|{\bar u}u|0>}\se 0.8\,, \eq
\bq  <D^{+}(p)|\,{\bar c}\,\Gamma_{\mu}\,d\cdot {\bar d}
\,\Gamma_{\nu}\,c\,|D^{+}(p)>_{\mu_o}\se p_{\mu}\,p_{\nu}\,f_D^2(\mu_o)+
\frac{2}{27}g_{\mu\,\nu}\,\rho_o\,r_D^2(\mu_o)\,.\eq
Therefore (see Eq.(55) in sect.2):
\bq  \Delta \Gamma^{lept}(D^{+})=\frac{1}{2\,M_D}<D^{+}(p)|\,\Delta
L^{lept}(\mu_o)\,|D^{+}(p)>_{\mu_o}\se  \nonumber  \eq
\bq  -\frac{G_F^2}{54 \pi\,M_D}\,M_D^2\,\rho_{o}\,
\tau_{co}\,r_D^2(\mu_o)\Psi_{o}\,, \quad
\Psi_{o}=\left [\frac{<{\bar s}s>^2}{<{\bar u}u>^2}
|V_{cs}|^2+|V_{cd}|^2 \right ]\se 0.656\,,  \eq
\bq  \frac{\Delta \Gamma^{lept}(D^{+})}
{\Gamma^{lept}_o}\se -\frac{32}{9}
\pi^2\frac{f_D^2(\mu_o)\,M_D^5}{M_c^7}\,\rho_{o}\,\tau_{co}\,
\Psi_{o}\se -0.045\,. \eq
In addition (see sects.3a,\,8 and fig.4)\,:
\bq  \delta \Gamma^{lept}_{PNV}\equiv \frac{1}{2\,M_D}
<D|\,L_{PNV}^{lept}(\mu_o)\,|D>_{\mu_o}= & \nonumber \\
& \displaystyle {{G_F^2}\over{12 \pi\,M_D}}
(1-\eta_{co}^{-2/9})\, \tau_{co}\,M_c^2
<D|\left [\,J^a_\rho(3)
\cdot {\bar c}{{\lambda^a}\over{2}}
\gamma_\mu\,(1+{{1}\over{3}}\gamma_5\,)\,c\,\right ]|D>_{\mu_o}\se
& \nonumber \\  & \displaystyle {{G_F^2}\over{12 \pi\,M_D}}(1-
\eta_{co}^{-2/9})\,\tau_{co}\,M_c^2 \left [
-{{4}\over{9}}\,r_D^2(\mu_o)\left (1-{{M_c^2}\over{2\,M_D^2}}
\right )\,\right ]\,, \eq
\bq  {{\delta \Gamma_{PNV}^{lept}}\over{\Gamma_{o}^{lept}}}\se - {{64}\over
{9}}\pi^2{{f_D^2(\mu_o)\,M_D^3}\over{M_c^5}}
(1-\eta_{co}^{-2/9})\,\tau_{co}\,
\left (1-{{M_c^2}\over{2\,M_D^2}}\right ) \se -0.085\,. \eq
On the whole:
\bq  \delta^{lept}(D^{+})\equiv {{1}\over{\Gamma_{sl}^o}}{{<D^+|\,\delta
L^{lept}_{eff}\,|D^+>}\over{2\,M_D}}
\se \left (-0.085-0.045 \right )\se -0.13\,. \eq

For the $B^{\pm,o,s}$ mesons (see sect.3b),\, the term $\Delta L^{lept}(\mu_o)$
gives no contribution,\, and the analog of eq.(198) looks as:
\bq  {{\delta \Gamma_{PNV}^{lept}}\over{\Gamma_{o}^{lept}}}\se  - {{16}\over
{3}}\pi^2{{f_B^2(\mu_o)\,M_B^3}\over{M_b^5}}
(1-\eta_{bc}^{-8/25})\,\tau_{bo}\,\eta_{co}^{-2/9}
\left (1-{{M_b^2}\over{2\,M_B^2}}\right ) \se -2\cdot 10^{-3}\,. \eq

All the above described contributions are nonvalence. There is also a
sizeable valence contribution to the $D_s$ leptonic width (see Eq.(56)
and sect.9):
\bq  \Delta \Gamma^{lept}(D_s)\se {{G_F^2}\over{4\pi\,M_D}}
|V_{cs}|^2\,(-0.245)\,
T_{\mu\,\nu}<D_s|{\bar c}\Gamma_\mu{{\lambda^a}\over{2}}
\,s\cdot {\bar s}\Gamma_{\nu}
{{\lambda^a}\over{2}}\,c|D_s>_{\mu_o}\se & \nonumber \\
& \displaystyle {{G_F^2}\over{4 \pi\,M_D}}|V_{cs}|^2\,
(-0.245)\,M_D^2 \left (\,C^c_G-P^c_V\, \right )\,, \eq
\bq  {{\Delta \Gamma^{lept}(D_s)}\over{\Gamma_{o}^{lept}}}
\se 48\pi^2|V_{cs}|^2
(-0.245){{M_D}\over{M_c^5}}\left (\,C_G^c-P_V^c\, \right )\se -7 \%\,. \eq \\

{\bf 12\,. \quad $\lambda$ and ${\bar \lambda}$\,.} \\

As it was indicated before,\, the values of $\lambda$ and ${\bar \lambda}$
(in terms of the c-quark and spectator quark momenta) are clear from each
diagram. Namely ($P_c$ is the c-quark
4-momuntum,\, $k_i$ are the spectator quark momenta,\, $P_D$ is the
D meson momentum,\, $P_D=P_{c1}+k_1=P_{c2}+k_2$ for the initial and final
D mesons):\\
a) $\,\lambda\se (P_c+k_1)=P_D$\, for the figs. 3a,\, 3b,\, 6a contributions;\\
b) $\,\lambda\se {\bar \lambda}\se P_D$\, (within logarithmic accuracy) for
the figs.9,\,10 contributions;\\
c) $\,\lambda\se {\bar \lambda}\se P_c\se M_c\,$ (within logarithmic accuracy)
for the fig.4 contribution;\\
d) $\,{\bar \lambda}\se ((P_D-k_1-k_2)\,$ for the figs.3c,\,6b contributions.

The differences between all of the cases above disappear in the formal limit
$M_Q\ra \infty\,$ but,\, as will be shown below,\, they are of great
importance for D mesons and are sizeable even for B mesons.

Because the D meson momentum,\, $P_D,\,$ is not an operator but a fixed
number,\, we have to deal practically with the case "d" only. So,\, let us
consider in detail the matrix element (see fig.3c and sect.3a):
\bq I_D\equiv <D(P_D)|\,{\bar \lambda}^2\,{\bar c}\Gamma_{\mu}\,d\cdot
{\bar d}\Gamma_{\mu}\,c\,|D(P_D)>\equiv <{\bar \lambda}^2_D>
<D|\,{\bar c}\Gamma_{\mu}\,d\cdot {\bar d}
\Gamma_{\mu}\,c\,|D>\,,\eq
where ${\bar \lambda}=(P_D-k_1-k_2)$ and $k_1,\,k_2\,$ are understood as
the 4-momentum operators of the initial and final spectator quarks.

As it was shown above (see sects.9,\,10),\, the factorization approximation
works very well even for the D mesons,\, and the non-factorizable
contributions are much smaller than factorizable ones. So,\, we can
reliably estimate $I_D$ as:
\bq I_D\se P^2_D <D|\,{\bar c}\,\Gamma_{\mu}\,d\,|0>
<0|\,{\bar d}\,\Gamma_{\mu}\,c\,|D>- & \nonumber \\
& \displaystyle 4 (P_D)_{\alpha} <D|\,{\bar c}\,\Gamma_{\mu}\,d\,|0>
<0|\,{\bar d}\,k_{\alpha}\Gamma_{\mu}\,c\,|D>
+2 <D|\,{\bar c}\,\Gamma_{\mu} k_{\alpha}\,d\,|0> <0|\,{\bar d}\, k_{\alpha}
\Gamma_{\mu}\,c\,|D>+ & \nonumber \\
& \displaystyle 2 <D|\,{\bar c}\,\Gamma_{\mu}\,d\,|0> <0|\,{\bar d}
\,k^2\Gamma_{\mu}\,c\,|D>\,. \eq
Let us define the matrix elements in Eq.(204) as:
\bq  <0|\,{\bar q}\,k_{\nu}i\gamma_5\,c\,|D(P)>=r_D\,<x>_{P}\,P_{\nu}\,,\eq
\bq  <0|\,{\bar q}\,k_{\nu}\gamma_{\mu}\gamma_5\,c\,|D(P)>=i\,f_D
<x>_{A}\left (\,P_{\mu}\,P_{\nu}-\frac{1}{4}\,g_{\mu\,\nu}
 \,P^2\,\right )\,,  \eq
\bq  <0|\,{\bar q}\,k^2\,\gamma_{\mu}\gamma_5\,c\,|D(P)>=i\,f_D\,P_{\mu}
\,<k^2>_A\,.\eq
The quantity $<x>$ has the meaning of the mean momentum fraction carried by the
light quark (in the $P_z\ra \infty\,$ frame),\, and $<k^2>$\, is the
characteristic value of the light quark 4-momentum squared
inside the D meson. So:
\bq  \frac{<{\bar \lambda}^2>}{M_D^2}\se \left (1-3<x>_A+\frac{3}{2}
<x>_A^2-2\,\frac{<-k^2>_A}{M_D^2} \right )\,.  \eq
Using the equations of motion for the matrix elements Eqs.(205),\,(206) it is
not difficult to obtain:
\bq  <x>_P=\frac{1}{2}\left (1-\frac{(M_c-m_q)^2}{M_D^2}\right ), \quad
<x>_A=\frac{4}{3}<x>_P \left [\,1+O\left ( \frac{\Lambda_{QCD}}{M_c}
\right )\right ].   \eq
As for the value of $<k^2>$,\, the estimates obtained from the corresponding
QCD sum rules for this quantity show that it is not far from its value
for the vacuum quarks (see Eq.(42)). So,\, we have
\footnote{That the effect is large for the D mesons can be seen from a
simplest rough estimate:
\bq k_1\se k_2\se <x>_A P_D\,, \quad {\bar \lambda}^2\se (P_D-2<x>_A P_D)^2
\se (1-2<x>_A)^2\,M_D^2\se 0.5\,M_D^2\,. \nonumber \eq}:
\bq <x>_A\se 0.15,\quad <-k^2>_A\se 0.4\,GeV^2,\, \quad
<{\bar \lambda}^{2}_{D}>\se 0.35\,M_D^2\,.\eq
It is seen that it is of great importance here to account for the spectator
quark momenta,\, in spite of the fact that these are power corrections
only in the formal limit $M_c \ra \infty$.
In what follows,\, we use the same estimate Eq.(210) also for the fig.6b
contribution:
\bq <D|\,\,{\bar \lambda}^2\,\,{\bar c}\,\Gamma_{\mu}\frac{\lambda^a}{2}
\,d\cdot {\bar d}\,\Gamma_{\mu}\frac{\lambda^a}{2}\,c\,|D>\se 0.35\,M_D^2
<D|\,{\bar c}\,\Gamma_{\mu}\frac{\lambda^a}{2}\,d\cdot {\bar d}
\,\Gamma_{\mu}\frac{\lambda^a}{2}\,c\,|D>\,.\eq

The corresponding expressions for the B mesons look as:
\bq <x>_A\se 6\%\,,\quad <-k^2>\se 0.4\,GeV^2,\,
\quad <{\bar \lambda}^{2}_{B}>\se 0.8\,M_B^2\,,\eq
so that the effect discussed is sizeable even for B mesons. \\

{\bf 13a\,. \quad Calculation of the $D^{\pm,o,s}$ decay widths}\\

Let us collect now the results obtained in previous sections (see also
Appendix). As it was shown in sects.3a,\,8,\, keeping only the leading
term and first corrections $O(1/M_c^2),\,$ we obtain a common nonleptonic
 decay width for all $D^{\pm,o,s}$ mesons:
\bq \Gamma_{nl}^o\se 1.59\,\Gamma_o\,, \quad \Gamma_o=\frac{G_F^2\,M_c^5}
{64\,\pi^3}\se 8.4\cdot 10^{-13}\,GeV\,.\eq
Accounting for the contributions of the four-fermion operators,\, we
obtained:\\

I)\,  The nonvalence contributions (which are the same for all
$D^{\pm,o,s}$ mesons (see sects.3a,10)) from the diagram of fig.4:
\bq \delta \Gamma_{PNV}\equiv{{1}\over{2\,M_D}}<D|\,L_{PNV}(\mu_o)\,|D>= &
\nonumber \\ & \displaystyle
{{G_F^2}\over{4 \pi\,M_D}}(1-\eta_{co}^{-2/9})\,M_c^2
 <D|\left [\,J^a_\rho(3)\cdot {\bar c}\,{{\lambda^a}\over{2}}
\gamma_\rho \left ( N_v+N_a\gamma_5\right )\,c\,\right ]|D>_
{\mu_o}\,\se & \nonumber \\
& \displaystyle {{G_F^2}\over{4 \pi\,M_D}}(1-\eta_{co}^{-2/9})\,M_c^2\,
\left [-{{4}\over{9}}\,r_D^2(\mu_o)\,N_v
\left (1-{{M_c^2}\over{2\,M_D^2}}\right )\,\right ]\,,\eq
\bq {{\delta \Gamma_{PNV}}\over{\Gamma_{o}}}\se - {{64}\over
{9}}\,\pi^2\,{{f_D^2(\mu_o)\,M_D^3}\over{M_c^5}}(1-\eta_{co}^{-2/9})\,N_v
\left (1-{{M_c^2}\over{2\,M_D^2}}\right ) \se -0.15\,.\eq

II)\, The nonvalence contribution from the diagram of fig.9 (see sect.10):
\bq \delta \Gamma_{NV}\equiv {{1}\over{2\,M_D}}<D|\Delta L(\mu_o)|D>_{NV}
\se {{G_F^2}\over{4 \pi\,M_D}}\,
{{2}\over{27}}\,M_D^2\,\rho_o\,\Omega_{NV}\,,& \nonumber \\
& \displaystyle \Omega_{NV}= \left [-A_u+A_d \left (|V_{ud}|^2+\xi_s
|V_{us}|^2\right )-
 A_s\left (\,\xi_s|V_{cs}|^2+|V_{cd}|^2
\right )\,\right ]\se -9.3\,, & \nonumber \\
& \displaystyle \xi_s=\frac{<{\bar s}s>^2}{<{\bar u}u>^2}\se 0.64\,,\eq
where the subscript NV means nonvalence contributions,\,
\bq {{\delta \Gamma_{NV}}\over{\Gamma_o}}\se - {{32}\over{27}}\,\pi^2\,{
{f_D^2(\mu_o)\,M_D^5}\over{M_c^7}}
\,\rho_o\,\Omega_{NV}\se -0.15\,.\eq

So,\, the total nonvalence contribution is:
\bq {{\Delta \Gamma_{NV}}\over{\Gamma_o}}\equiv {{[\delta \Gamma_{PNV}]+
[\delta \Gamma_{NV}]}\over{\Gamma_{o}}}\se \left [-15\%\,\right ]+
\left [\,-15\%\,\right ]=-30\%\,.\eq

III)\, We have for the valence contributions (see sects.3a,\,9,\,10 and
figs.6,\,10):
\bq \Delta \Gamma_{nl}(D^o)\equiv \frac{1}{2\,M_D}<D^o|\Delta L^{(c)}(\mu_o)
|D^o>_{V}\se \frac{G_F^2}{4\pi\,M_D}\,O_u\,M_D^2\,
\left (\,C_G^c-P_V^c\,\right )\,,\eq
\bq {{\Delta \Gamma_{nl}(D^o)}\over{\Gamma_o}}
\se 16\,\pi^2\,{{M_D}\over{M_c^5}}\,O_u
\left (\,C_G^c-P_V^c\, \right )\se +33\%\,.\eq

\bq {{\Delta \Gamma_{nl}(D^+)}\over{\Gamma_o}}={{1}\over{\Gamma_o}}
<D^+|\Delta L^{(c)}(\mu_o)|D^+>_{V}
\se \left [\,16\,\pi^2\,{{f_D^2(\mu_o)\,M_D^3}\over{M_c^5}}
|V_{ud}|^2\,S_d\,\frac{<{\bar \lambda}^2_D>}{M_D^2}\,
\right ]- & \nonumber \\
& \displaystyle \left [\, 48\,\pi^2\,{{M_D}\over{M_c^5}}
|V_{ud}|^2\,O_d\left ({{{<\bar
\lambda}^2_D>}\over{M_D^2}}\,C_G^c-{{4}\over{3}}\,P_V^c\, \right )\right ]
\se \left [\,-62\%\right ]+ \left [\,-21\%\right ]\se -83\%\,,\eq
where the subscript V means valence contributions.

For the $D_s$ meson \footnote{\,I am indebted to
N.G.\,Uraltsev who pointed out an arithmetical mistake in the original
calculation of this correction.}:
\bq \frac{\Delta \Gamma_{nl}(D_s)}{\Gamma_o}=\frac{1}{\Gamma_o}
<D_s|\Delta L^{(c)}(\mu_o)|D_s>_{V}\se \left [ 16\,\pi^2\,|V_{us}|^2
\frac{f_D^2(\mu_o)\,M_D^3}{M_c^5}\,S_d\,\frac{<{\bar \lambda}^2_D>}
{M_D^2}\right ]+ & \nonumber \\
& \displaystyle \left [\,16\,\pi^2\,|V_{cs}|^2\,\frac{M_D}{M_c^5}\,O_s
\left ( \,C^c_G-P^c_V\,\right )\,\right ]+ & \nonumber \\
& \displaystyle \left [\,-48\,\pi^2\,|V_{us}|^2\,\frac{M_D}{M_c^5}\,O_d
\left (\,\frac{<{\bar \lambda}^2_D>}{M_D^2}\,C_G^c-\frac{4}{3}\,P_V^c\,
\right )\,\right ]\se & \nonumber \\
& \displaystyle \left [\,-3\%\,\right ]+\left [\,-3\%\,\right ]
+\left [\,-1\%\,\right ]\se -7\%\,. \eq

We have therefore for the nonleptonic widths (the experimental values are
given in brackets,\, all values are given below in units of $10^{-13}\,GeV$):
\bq \Gamma_{nl}(D^+)\se \Gamma_o\left [\,1.59-0.30-0.83\,\right ]\se
0.46\,\Gamma_o\se 3.9 \quad \left \{ \, 4.05\, \right \}\,,\eq
\bq \Gamma_{nl}(D^o)\se \Gamma_o\left [\,1.59-0.30+0.33\,\right ]\se
1.62\,\Gamma_o\se 13.6 \quad \left \{ \,13.5\, \right \}\,,\eq
\bq \Gamma_{nl}(D_s)\se \Gamma_o\left [\,1.59-0.30-0.07\,\right ]\se
1.22\,\Gamma_o\se 10.2 \quad \left \{\,\,\, ? \,\,\,\right \}\,,\eq
We have for the semileptonic widths (see sects.4,\,11):
\bq \Gamma^{lept}(D^+)=1.06 \quad (input) \,,\eq
\bq \Gamma^{lept}(D^o)\se  \Gamma^{lept}(D^+)\,,\eq
\bq {\tilde \Gamma}^{lept}(D_s)\se \Gamma^{lept}(D^+)\left [\,1-7\%\,\right ]
\se 1.0\,\,. \eq
We have to add also to $\Gamma^{lept}(D_s)$ the $"D_s\ra \tau\,\nu"$
contribution:
\bq \frac{\Gamma(D_s\ra \tau\,\nu)}{\Gamma^{lept}_o}=24\,\pi^2\frac{f^2_{D_s}
(M_D)\,M_{D_s}\,M^2_{\tau}}{M_c^5}\left [1-\frac{M^2_{\tau}}
{M^2_{D_s}}\right ]^2\se 0.16\left (\frac{f_{D_s}(M_D)}{200\,MeV}\right )^2\,,
\nonumber & \\ & \displaystyle \Gamma(D_s\ra \tau\,\nu)\se 0.4\,,\quad
\Gamma^{lept}_{tot}(D_s)\se 2\cdot 1+0.4\se 2.4\,.\eq
So,\, we have finally for the total decay widths:
\bq \Gamma_{tot}(D^+)\se \left [\,3.9+2.1\,\right ] \se 6.0\,, \quad
\left \{\,6.2\,\right \}\eq
\bq \Gamma_{tot}(D^o)\se \left [\,13.6+2.1\,\right ] \se 15.7\,,
\quad \left \{\,15.6 \,\right \}\eq
\bq \Gamma_{tot}(D_s)\se \left [\,10.2+2.4\,\right ] \se 12.6\,,
\quad \left \{\,13.8\,\right \}\,.\eq \\

{\bf 13b\,. \quad Calculation of the $B^{\pm,o,s}$ lifetime
differences}\\

Substituting the corresponding numbers into the formula Eq.(25), we obtain:
\bq \frac{2\,C_{+}^2+C_{-}^2}{3}\se 1.127\,,\quad \frac{<B|\,{\bar b}\,b\,|B>}
{2\,M_B}\se 1.00\,,\quad I_{rad}\se 0.98\,,\eq
\bq \Gamma^o_{nl}\se 1.10\,\Gamma_o\,z_o\,, \quad \Gamma_o=\frac
{G_F^2\,M_b^5\,|V_{cb}|^2}{64\,\pi^3}\,.\eq
The values of the phase space factors $z^{i}_{o}$ \cite{Cor}
(for $M_b=5.04\,GeV,\, M_c=1.65\,GeV,\, M_{\tau}=1.784\,GeV $) are:
\bq  z_{o}^{cud}\se 0.460\,,\quad z_{o}^{ccs}\se 0.131\,,
\quad z_{o}^{c\tau\nu} \se 0.105\,.\eq
So,\, neglecting the 4-fermion operator contributions (which are small,\,
see below) we have for the relative yields:
\bq n^{eff}_B\se \biggl \{\,(1)_{ud}+(0.285)_{cs}+2\,(0.260)_{e\nu+\mu\nu}+
(0.060)_{\tau\nu}\,\biggr \}\se 1.87\,,\eq
and the branching fractions are respectively:
\bq Br\,(b\ra c{\bar c}s)\se 15\%\,, \quad  Br\,(b\ra c e\nu)\se 13.9\%\,,
 & \nonumber \\ & \displaystyle
Br\,(b\ra c\tau\nu))\se 3.2\%\,, \quad Br\,(\frac{\tau}{e})\se 0.23\,.\eq
The above value of $Br\,(b\ra c e\nu)$ is $\se 20\%$ larger than the LEP data
\cite{MD},\, and this is a well known difficulty \cite{A2},\,\cite{PS},\,
\cite{BBSV}. What we can add is that
the 4-fermion operator contributions calculated below are too small
and can not help here. From our viewpoint,\, one of the important reasons
for this discrepancy is that the radiative corrections to the nonleptonic
widths are known for massless quarks only,\, and there are reasons to
expect that accounting for nonzero final quark masses will increase
considerably the nonleptonic radiative correction.
\footnote{ To illustrate possible
changes,\, we can put $M_c=0$ in the leptonic radiative correction and
obtain: $Br\,(b\ra c e\nu)\se 12.9\%\,,$ in comparison with $\se 11.4\%$ from
the LEP data.} Another evident reason which comes to mind is that the energy
release is not large in the $b\ra c{\bar c}s$ mode,\, so that it can be
enhanced due to nonperturbative effects. Although the present data do not
support this,\, it seems that the experimental numbers can
change here with time.

Let us proceed now to the calculation of the B meson lifetime differences.
Using results obtained in preceeding sections (see sects.\,3b,9,10)
,\, we have for the valence contributions:
\bq {{\Delta \Gamma(B^-)}\over{\Gamma_o}}\se 16\pi^2\,
{{<{\bar \lambda}^{2}_{B}>}\over
{M_B^2}}\left (1-{{M_c^2}\over{<{\bar \lambda}^{2}_{B}>}}\right )^2 \times
 & \nonumber \\ & \displaystyle
\left \{S_d {{f_B^2(\mu_o)\,M_B^3}\over{M_b^5}}-3\,O_d\,{{M_B}
\over{M_b^5}}\left (C_G^B-{{4}\over{3}}\xi_B\,P_V^B\right )\right \}\se &
\nonumber \\ & \displaystyle
\left (\,-1.9\%\,\right )+\left (-1.3\%\,\right )
\se -3.2\%,\eq
\bq \xi_B={{M_B^2}\over{<{\bar \lambda}^{2}_{B}>}}\,{{\left (1-{{M_c^2}\over
{M_B^2}}\right )^2}\over{\left (1-{{M_c^2}\over{<{\bar \lambda}^{2}_{B}>}}
\right )^2}}\se 1.31\,.\eq
\bq \frac{\Delta \Gamma(B^o)}{\Gamma_o}\se 16\pi^2\,(1-x)^2 \frac{M_B}
{M_b^5}\,O_u\,\left [(1+\frac{x}{2})\,C_G^B-P_V^B\,\right]\,\se 0.6\%\,,\eq
\bq \frac{\Delta \Gamma(B_s)}{\Gamma_o}
\se 16\pi^2\,(1-4\,x)^{1/2}\,\frac{M_B}
{M_b^5}\,O_u\left [\,(1-x)\,C_G^B-(1-2\,x)\,P_V^B\, \right ]\se 0.5\%\,,\eq
where: $x=M_c^2/M_B^2$\,.
The nonvalence penguin contributions are also small,\,
both factorizable (see fig.4) and nonfactorizable (see fig.9):
\bq \frac{\Delta \Gamma_{PNV}}{\Gamma_o}\se -\frac{64}{9}\,\pi^2\,
\frac{f_B^2(\mu_o)\,M_B^3}{M_b^5}\,\left (\,1-
\frac{M_b^2}{2\,M_B^2}\,\right )\cdot A\se -0.5\%\,.\eq
\bq \frac{\Delta \Gamma_{NV}}{\Gamma_o}\se -
\frac{32}{27}\pi^2\frac{f_b^2(\mu_o)\,M_B^5}{M_b^7}
\,\rho_{o} \times  & \nonumber \\
 & \displaystyle \left [\,4(1-x)^2
A_d-A_u-\sqrt{1-4x}\,(1-2\,x)\frac{<{\bar s}s>^2}{<{\bar u}u>^2}A_u\,\right ]
\se -0.1\%\,.\eq
It is seen that,\, analogously to the D mesons,\, the largest effect
is the negative contribution to the $B^-$ width due to Pauli interference,\,
but it is only $\se -3\%$ here,\, while the $B^o$ and $B_s$ meson widths
receive both only $\se 0.5\%$ corrections.\\

Let us comment finally on the accuracy of all the above calculations.
It is extremely difficult to give a reliable estimate of the accuracy. Too
many issues are involved,\, and estimates of various contributions vary in
their accuracy. So,\, the accuracy of predictions for various quantities
(the quark masses,\, $|V_{cb}|$,\, the D meson decay widths,\, the B meson
lifetime differences,\, etc.) differ from each other considerably.
So,\, we do not even try to give here any concrete numbers,\, except
for some "educated guesses" like: a) the c- and b-quark masses can
hardly be less than 1.6\,GeV and 5.0\,GeV respectively; b) $f_B(M_b)$ can
hardly be larger than 120\,MeV; c) the value of
$|V_{cb}|$ can hardly deviate more than $\pm 0.002$ from 0.040 (with the
experimental value of the semileptonic width taken at its central value,\,
see Eq.(114)); d) the lifetime difference between the $B^-$ and $B^o$ mesons
can hardly be more than $\se 5\%$.\\

{\bf 14\,. \quad $B^o-{\bar B}^o$ mixing} \\

The effective Lagrangian which determines the mixing width,\,
$\Gamma_{12},\,$ is obtained directly
from Eqs.(61),(65) by the substitution: $({\bar b}...s)\ra ({\bar s}...b)$
\footnote{ It is implied in Eq.(244) that when calculating the matrix
element each of two b-operators acts on both sides,\, and this gives the
additional factor 2 which is compensated by the additional factor 1/2
introduced into Eq.(244)}:
\bq L_{width}^{(mix)}(M_b)=\frac{G_F^2\,(V_{cb}V_{cq}^{*})^2}{4\pi}
T^{({\bar c}c)}_{\mu\,\nu}\,L^{(mix)}_{\mu\,\nu}(M_b)\,, & \nonumber \\
& \displaystyle L^{(mix)}_{\mu\,\nu}(M_b)=S_u^o\,
(\,{\bar s}\,\Gamma_{\mu}\,b\,)
\,(\,{\bar s}\,\Gamma_{\nu}\,b\,)+O_u^o\,(\,{\bar s}\frac{\lambda^a}{2}
\Gamma_{\mu}\,b\,)\,(\,{\bar s}\frac{\lambda^a}{2}\Gamma_{\nu}\,b\,)\,,\eq
where q denotes the s or d quark field.
Because the difference between $\lambda$ and $\bar \lambda$ (see sect.12) is
not of great importance for the B mesons but complicates considerably the
renormalization formulae,\, it will be neglected in what follows. Then,\,
using the relation:
\bq (\,{\bar s}\frac{\lambda^a}{2}
\Gamma_{\mu}\,b\,)\,(\,{\bar s}\frac{\lambda^a}{2}\Gamma_{\nu}\,b\,)=-
\frac{2}{3}(\,{\bar s}\,\Gamma_{\mu}\,b\,)
\,(\,{\bar s}\,\Gamma_{\nu}\,b\,)+\frac{1}{4} \,g_{\mu\,\nu}(\,{\bar s}\,
\Gamma_{\rho}\,b\,)\left (\,{\bar s}\,\Gamma_{\rho}\,b\,\right )\,, \eq
we can rewrite Eq.(244) in the form:
\bq L^{(mix)}_{\mu\,\nu}(M_b)=\left (S_u^o-\frac{2}{3}O_u^o\right )\left
[\,{\bar s}\,\Gamma_{\mu}\,b\cdot {\bar s}\,\Gamma_{\nu}\,b-\frac{1}{4}\,
g_{\mu\,\nu}\,{\bar s}\,\Gamma_{\rho}\,b\cdot {\bar s}\,\Gamma_{\rho}\,b\,
\right ]_{M_b}+  & \nonumber \\ & \displaystyle
\frac{1}{4}\left (S_u^o+\frac{1}{3}O_u^o\right )\,g_{\mu\,\nu}
\left [\,{\bar s}\,\Gamma_{\rho}\,b\cdot {\bar s}\,\Gamma_{\rho}\,b\,\right ]
_{M_b}\,.\eq
The operators in the square brackets in Eq.(246) renormalize multiplicatively
\cite{VUKS},\, so that we obtain:
\bq L^{(mix)}_{\mu\,\nu}(\mu_o)=\alpha \left [\,{\bar s}\,\Gamma_{\mu}\,b
\cdot {\bar s}\,\Gamma_{\nu}\,b\,\right ]_{\mu_o}+\beta\left [\,
g_{\mu\,\nu}\,{\bar s}\,\Gamma_{\rho}\,b\cdot {\bar s}\,\Gamma_{\rho}\,b\,
\right ]_{\mu_o}\,, & \nonumber \\ & \displaystyle
\alpha=\Lambda_{bo}^{1/3}\left (S_u^o-\frac{2}{3}O_u^o\right )\se -1.93\,,
& \nonumber \\ & \displaystyle
\beta=\frac{1}{4}\left [\Lambda_{bo}\left (\frac{1}{3}O_u^o+S_u^o
\right )+\Lambda_{bo}^{1/3}\left (\frac{2}{3}O_u^o-S_u^o\right )\right ]
\se 0.83\,, & \nonumber \\ & \displaystyle
\Lambda_{bo}=\left (\frac{\alpha_s(M_c)}{\alpha_s(M_b)}\right )^{12/25}\left
(\frac{\alpha_s(\mu_o)}{\alpha_s(M_c)}\right )^{4/9}\se 1.614\,\,.\eq
Using now (see sects.8,\,9):
\bq <{\bar B}^o(p)|\,L^{(mix)}_{\mu\,\nu}(\mu_o)\,|B^o(p)>_{factor}\se
2\,\alpha\, f_B^2(\mu_o) \left (\frac{2}{3}\,p_{\mu}\,p_{\nu}+
\frac{1}{6}\,p^2\,g_{\mu\,\nu}
\right )+ & \nonumber \\ & \displaystyle
2\,\beta\,\frac{4}{3}f_B^2(\mu_o)\,M_B^2\,g_{\mu\,\nu}\,, & \nonumber \\
 & \displaystyle <{\bar B}^o(p)|\,L^{(mix)}_{\mu\,\nu}(\mu_o)\,|B^o(p)>_{
nonfactor}\se 2 \left [-2\,\alpha\,\frac{p_\mu\,p_\nu}{p^2}\,C_G^b\right ]+
& \nonumber \\ & \displaystyle
2\,g_{\mu\,\nu}\Bigl [\,\alpha\left (-C_G^b+2\,P_V^b\right )+\beta
\left (-6\,C_G^b+8\,P_V^b\right )\Bigr ] \,,\eq
and contracting with:
\bq T_{\mu\,\nu}^{({\bar c}c)}=\sqrt {1-4\,x}\left \{\frac{1+2\,x}{3}\left (
p_\mu\,p_\nu-p^2\,g_{\mu\,\nu}\right )+p^2\,x\,g_{\mu\,\nu}\right \}\,,\quad
x\se \frac{M_c^2}{M_b^2}\,, \eq
we have:
\bq T_{\mu\,\nu}^{({\bar c}c)}<{\bar B}^o(p)|\,L^{(mix)}_{\mu\,\nu}(\mu_o)\,|
B^o(p)>_{factor}\se 2\,(-0.57)\,f_B^2(\mu_o)\,M_B^4\se -7.0\,GeV^6\,,
& \nonumber\\  & \displaystyle
T_{\mu\,\nu}^{({\bar c}c)}<{\bar B}^o(p)|\,L^{(mix)}_{\mu\,\nu}(\mu_o)\,|
B^o(p)>_{nonfactor}\se 1.2\,GeV^6\,. \eq
So,\, we have finally for $\Gamma_{12}$:
\bq \frac{\Gamma_{12}(B)}{\Gamma_o}=\frac{1}{\Gamma_o}\frac{
<{\bar B}^o|\,L_{width}^{(mix)}\,|B^o>}{2\,M_B}\se & \nonumber \\
 & \displaystyle \xi_{KM}\,16\pi^2\frac
{f_B^2(\mu_o)\,M_B^3}{M_b^5}\left [\,-0.57\left (1-17\%\right )\right ]
\se (-2.7\%)\,\xi_{KM}\,,\eq
where $\xi_{KM}$ is the ratio of the Kobayashi-Maskawa factors.
It is seen that the nonfactorizable contributions appear to be surprisingly
large here and decrease the mixing by $\se 17\%$.

An analogous situation takes place for the mixing mass of the $B^o$ and
${\bar B}^o$ mesons. The effective Lagrangian can be found in any review
and has the form:
\bq L_{mass}^{(mix)}(M_b)=C_o\left [{\bar s}\,\Gamma_{\rho}\,b\cdot {\bar s}\,
\Gamma_{\rho}\,b\,\right ]_{M_b}\,,\eq
where $C_o$ is a known coefficient (see sect.15). The above operator
renormalizes multiplicatively \cite{VSG},\, so that there appears only the
factor $\Lambda_{bo}$ (see Eq.(247)) when it is renormalized to the point
$\mu_o$. The matrix element is given by Eq.(248),\, so that we have:
\bq <{\bar B}^o|\,L_{mass}^{(mix)}(M_b)\,|B^o>_{factor}\se \Lambda_{bo}\,
\frac{8}{3}f_B^2(\mu_o) M_B^2\,C_o\se 0.95\,GeV^4\,C_o\,, & \nonumber \\
& \displaystyle <{\bar B}^o|\,L_{mass}^{(mix)}(M_b)\,|B^o>_{nonfactor}\se 2
\Lambda_{bo}\left [-6 C_G^b+8 P_V^b\right ] C_o\se -0.176\,GeV^4\,C_o\,.\eq
On the whole:
\bq <{\bar B}^o|\,L_{mass}^{(mix)}(M_b)\,|B^o>_{M_b}\se \Lambda_{bo}\,
\frac{8}{3}f_B^2(\mu_o) M_B^2\left (\,1-18\%\,\right )C_o\se & \nonumber \\
 & \displaystyle \frac{8}{3}f_B^2(M_b) M_B^2\left (\,1-18\%\,\right ) C_o\se
0.8\,GeV^4\,C_o\,;\,\,\, B_B(M_b)\se (1-0.18)\se 0.82\,,\eq
and the corrections to the factorization approximation are also very
significant here. \\

{\bf 15\,. \quad The unitarity triangle}\\

The purpose of this section is to show that the results obtained above are
marginally consistent with the available data and can be used to determine
the parameters of the unitarity triangle (we use below in this section the
Wolfenstein parametrization and the notations from \cite{BH}).

We would like to emphasize that we have not tried in this section to account
for all the various predictions available in the literature for
the parameters involved
($f_B,\, B_B,\, B_K,\,$ etc.). In fact,\, there are a number of papers which
try to account with an equal weight for all the values available in the
literature for these parameters. This results in large uncertainties which
prevent to obtain more or less definite results from the
available experimental data. Instead,\, we prefer to use only those results
which,\, from our viewpoint,\, are more reliable and this allows us to obtain
the concrete predictions for the unitarity triangle parameters. \\

{\bf 1)\,.}  Using: $V_{cb}=A\,\lambda^2\,,\quad \lambda\se 0.22$\,\, and (see
Eq.(114))\, $V_{cb}\se 0.040$\,, one has:
\bq A \se 0.825\,. \eq

{\bf 2)\,.} The $B_d-{\bar B}_d$ mass difference is given by:
\bq x_d\equiv \frac{\Delta M_d}{\Gamma_B}\se \tau_B\,\frac{G_F^2}{6\pi^2}\,
M_W^2\,M_B\left [f_B^2\,B_B\right ]_{M_B}\left [{\bar \eta}^*_{2B}\,
S(x^*_t)\right ]\,|V_{td}|^2 =(0.71\pm 0.07)\,,\eq
\bq S(x)=x\left [\frac{1}{4}-\frac{9}{4(x-1)}-\frac{3}{2(x-1)^2}\right ]+
\frac{3}{2}\left (\frac{x}{x-1}\right )^3\log x\,,\nonumber \eq
\bq x^*_t=\left (\frac{M^*_t}{M_W}\right )^2\,,\,\,\, |V_{td}|^2=A^2\,
\lambda^6\left [\,(1-\rho)^2+\eta^2\right ]\se
0.77\cdot 10^{-4}\left [\,(1-\rho)^2+\eta^2\right ]\,.\eq
Using in Eq.(256) ( see Eqs.(128),\,(254) and \cite{BH} )
\footnote {In order to use the numericals from \cite{BH} we use the mass
$M^*_t$ which is the so-called ${\overline {MS}}$ mass;\, $M^*_t=180\,GeV$
corresponds to the pole mass:\,\,$M^{pole}_t\se 190\,GeV$\,,\, which is
marginally consistent with the CDF and LEP data,\, \cite{CDF}.}:
\bq M^*_t\se 180\,GeV\,,\quad \left [{\bar \eta}^*_{2B}\,
S(x^*_t)\right ] \se 0.85 \cdot 2.7 \se 2.30\,, \eq
\bq \tau_B\se 1.6\cdot 10^{-12}\,s\,,\quad f_B(M_b)\se 113\,MeV\,,\quad
B_B(M_b)\se 0.82\,,\eq
one obtains:
\bq \left [\,(1-\rho)^2+\eta^2\right ]\se \frac{x_d}{0.35}\se 2.0\,.\eq

{\bf 3)\,.} Using also:
\bq \left |\frac{V_{ub}}{V_{cb}}\right |^2=\lambda^2\left (\rho^2+\eta^2
\right )\se 1\cdot 10^{-2}\,,\quad \left (\rho^2+\eta^2\right )\se 0.2\,,\eq
one obtains then from Eqs.(260),\,(261):
\bq \rho\se -0.4 \,,\quad \eta\se 0.2 \,,\quad  \delta=tg^{-1}(\frac{\eta}
{\rho})\se 0.85\,\pi\,,\eq
\bq \sin\,(2\,\alpha)\se 0.60\,,\quad \sin\,(2\,\beta)\se 0.28\,,\quad
\sin\,(2\,\gamma)\se -0.80\,.\eq

{\bf 4)\,.} The CP-violating part of the $K^{o}-{\bar K}^{o}$ mixing can be
written in the form\, \cite{BH}:
\bq e^{-i\pi/4}\,\epsilon_K\se C_{\epsilon}\,B_K\,A^2\,\lambda^6\,\eta
\left [P_o+A^2\,\lambda^4\,(1-\rho\,)
\eta^*_{2K}\,S(x^*_t)\right ]\,,\eq
\bq C_{\epsilon}=\frac{G_F^2\,F_K^2\,M_K\,M_W^2}{6\sqrt{2}\,\pi^2\,\Delta M_K}
\se 3.85\cdot 10^4\,,\eq
\bq P_o=x_c\left (\,\eta_3\,S_3(x_t)-\eta_1\,\right )\,,\quad
 S_3(x_t)=\left [ \log \frac{x_t}{x_c}-\frac{3}{4}\frac{x_t}{x_t-1}\left (
\frac{x_t\,\log {x_t}}{x_t-1}-1\right )\right ]\,,\eq
\bq x_i=\left ( \frac{M_i}{M_W}\right )^2\,,\quad M_c\se 1.65\,GeV\,,
\quad M^*_t\se 180\,GeV\,,\eq
\bq \eta_1\se 0.80\,,\quad \eta^*_{2K}\se 0.57\,,\quad \eta_3\se 0.36\,. \eq
Substituting all this into Eq.(264),\, one has:
\bq e^{-i\pi/4}\,\epsilon_K=2.26\cdot 10^{-3}\se 7.3\cdot 10^{-3}\,B_K\,
\eta \left [\,(1-\rho)+\omega_o\,\right ]\,,\eq
\bq \omega_o=\frac{P_o}{|V_{cb}|^2\,\eta^*_{2K}\,S(x^*_t)}
\se 0.38\,,\nonumber \eq
\bq \eta\left (1-\rho+\omega_o\right )\se \frac{0.31}{B_K}\,.\eq
Let us recall\, \cite{BH} that,\, unlike $B_B(M_b)$ in Eq.(254),\, $B_K$ in
Eq.(264) is defined as:
\bq <{\bar K}^o|\,{\bar s}\,\Gamma_{\nu}\,d\cdot {\bar s}\,\Gamma_{\nu}\,d\,
|K^o>_{\mu}\equiv \frac{8}{3}\,B_K(\mu)\,f_K^2\,M_K^2\,,\eq
\bq B_K=B_K(\mu)\,\alpha_{s}^{-2/9}(\mu)\,.\eq
The characteristic value of $B_K$ obtained from lattice calculations is:
$B_K=(0.9\pm 0.1)$\, \cite{CB}\,, and substituting $B_K\se 0.9$ into
Eq.(270),\, one obtains:
\bq \eta\left (1-\rho+\omega_o \right )\se 0.345\,,\eq
Using now: $\rho\se -0.40,\,\, \omega_o\se 0.38\,$
one obtains $\eta\se 0.19,\,$ in agreement with Eq.(262).

{\bf 5)\,.} Clearly,\, we can proceed in the opposite way:
using $B_K\se 0.9$ we obtain from $\epsilon_K$ and $x_d$:
\bq \eta\,\left (1-\rho+0.38\,\right )\se 0.345\,,\quad (1-\rho)^2+\eta^2\se
2.0\,\,.\eq
This gives us then: $\,\rho\se -0.4\,,\,\,\eta\se 0.2\,\,$ and
$\,|V_{ub}/V_{cb}|\se 0.10\,.$

With the above parameters,\, the CP-violating asymmetry in $B^o\ra
\Psi\,K_S$ decay is:
\bq A(B^o\ra \Psi\,K_S)\se \frac{x_d}{1+x_d^2}\sin (2\,\beta)\se 0.13\,.
\nonumber \eq

{\bf 6)\,.} The quantity $\epsilon ^{\prime}/\epsilon$\, can be
represented in the form \cite{BJL}:
\bq \frac{\epsilon ^{\prime}}{\epsilon}=\frac{Im\,\lambda_t}{1.7}\left [\,P^
{(1/2)}-P^{(3/2)}\right ]\se 0.4\cdot 10^{-4}\left [\,P^{(1/2)}-P^{(3/2)}
\right ]\,,\eq
\bq Im\,\lambda_t=\eta\,A^2\,\lambda^5\se 0.7\cdot 10^{-4}\,,\eq
where $P^{(1/2)}$ and $P^{(3/2)}$ are expressed through the known coefficients
$a_i(M_t)$ and the parameters $B_6^{(1/2)},\,B_8^{(3/2)}$. These latter
are determined from the
corresponding matrix elements of the penguin operators,\, see \cite{BJL}.
In the factorization approximation: $B_6^{(1/2)}=B_8^{(3/2)}\equiv 1$,\, and
the values obtained from both the lattice calculations and the
$1/N_c$ expansion agree with the factorization approximation within
$\pm 20\%$.
Using the values of $a_i(LO)$ from the Tables 14,\,15 in \cite{BJL} and
$B_6^{(1/2)}\se B_8^{(3/2)}\se 1$,\, one obtains:
\bq  P^{(1/2)}\se 5.2\,,\quad P^{(3/2)}\se 6.2\,,\quad
\frac{\epsilon ^{\prime}}{\epsilon}\se -0.4\cdot 10^{-4}\,.\eq
The above value can not be taken too literally because it is a result of
strong cancelations. \footnote{These are due to the large value of
the t-quark mass which enhances the electroweak penguin contribution
$P^{(3/2)}$,\, see \cite{Flynn}\,.}
To be conservative, the realistic value of
$\epsilon ^{\prime}/\epsilon$ can be considered to lie in the
interval: $\sim \pm (1-2)\cdot 10^{-4}$. In any case,\, this is inconsistent
with the NA31-result:
$\epsilon ^{\prime}/\epsilon=(\,23\pm 7\,)\cdot 10^{-4}$,\,
while there is no contradiction with the E731-result:
$\epsilon ^{\prime}/\epsilon=(\,7\pm 6\,)\cdot 10^{-4},\,$
\cite{Bur},\,\cite{Pat},\,\cite{BW}.

{\bf 7)\,.} The box diagram contribution to $(K_L-K_S)$ the mass difference,\,
$\Delta M_K=M(K_L)-M(K_S)$,\, is \cite{Vys},\,\cite{GW}:
\bq \left \{ \Delta M_K \right \}_{box}=\frac{G_F^2}{6\pi^2}\,f_K^2\,M_K
\,M_c^2\,\lambda^2\,B_K\,\eta_1 \left [\,1+O(10^{-2})\,\right ]\,,\eq
where the correction $O(10^{-2})$ represents the
t-quark contribution and other small corrections.
Using: $f_K\se 162\,MeV,\, M_c\se 1.65\,GeV,\, B_K\se
0.9$,\, one obtains:
\bq \left \{ \Delta M_K \right \}_{box}\se \left (4.0\cdot 10^{-15}\,GeV
\right )B_K\,\eta_1 & \nonumber \\ & \displaystyle
\se \left (3.6\cdot 10^{-15}\,GeV \right )\eta_1 \se 2.9\cdot
10^{-15}\,GeV\,, \eq
for $\,\eta_1\se 0.8$.\, The experimental value is: $\Delta M_K=3.5
\cdot 10^{-15}\,GeV,\,$ so that there is not much room for additional
large distance contributions. \\

{\bf 16\,. \quad Summary and conclusions} \\

Let us summarize first in short the results obtained for the D mesons.

It is seen that the
overall picture is sufficiently complicated: there are a number of
contributions, all of the same order and of different signs. As to
the four-fermion operator contributions we were mainly interested in,\,
the qualitative picture is as follows.

Although they are formally $0(1/M_c^3)$ corrections,\,
these contributions
are very important numerically and comparable with the Born term.
In particular,\, their final effect in
the nonleptonic widths is much larger than
those of the leading $0(1/M_c^2)$ corrections. There are two reasons for this.

{\bf i}) Various $0(1/M_c^2)$ corrections,\, being only a few times smaller
than the Born term,\, cancel each other strongly,\, see Eq.(147);

{\bf ii}) The four-fermion operator contributions are the first to gain a
large numerical factor from the larger two-particle phase space. Otherwise
they would have been much smaller than separate $0(1/M_c^2)$ terms.
It is clear that this effect operates one time only. So, there are reasons to
expect that all other $O(1/M^3)$ and higher order
corrections are much smaller,\, and those considered in this paper are the
main ones. Some simple estimates confirm this.

As for the relative significance of various contributions,\, the picture is
as follows.

{\bf 1})\, Most significant is the destructive Pauli interference effect
(see fig.3c) of two d-quarks in $D^-$ decay ($\sim -60\%$ of the Born term,\,
$\Gamma_o\,$).

{\bf 2})\, The (cross) annihilation contribution (see fig.6b) ensured by the
nonperturbative nonfactorizable gluon interaction decreases
further ($\sim - 20\%$ of $\Gamma_o$) the $D^-$ nonleptonic width.

{\bf 3}) Both the above contributions could have appeared much larger,\,
but are strongly suppressed by the much smaller two particle phase space. The
reason is that the "normal" total 4-momentum of the quark pair is
(see fig.3a): $\lambda\se (P_c+k_1)=P_D$ (k is the momentum of the spectator
quark). However,\, it is: ${\bar \lambda}\se (P_c-k_2)=(P_D-k_1-k_2)$ for both
these "cross-contributions". Because the charm quark is not very heavy and
spectator quarks carry a significant fraction of the charmed meson momentum,
\, this leads to a strong suppression: $\lambda^2\gg {\bar \lambda}^2\,.$
This effect remains noticeable even for the B mesons.

{\bf 4})\, The (direct) weak annihilation contribution (see fig.6a) increases
significantly ($\sim 30\%$ of $\Gamma_o\,$) the $D^o$ nonleptonic width
\footnote{Let us emphasize that the annihilation contribution $\se 30\%$ into
the inclusive width does not contradict that it can be $\sim 100\%$ in separate
exclusive modes,\, see \cite{C0}.}.

{\bf 5})\, The nonvalence penguin contributions,\, both factorizable (see
fig.4) and nonfactorizable (see fig.9)),\, are very significant and,\, on the
whole,\, diminish considerably (\,$\sim -(20-25)\%\,\,$) both semileptonic and
nonleptonic widths.

{\bf 6})\, There are no noticeable positive valence contributions into the
$D_s$ meson width. Those which are available are negative and decrease $(\sim
-7\%$ of the Born terms) both semileptonic and nonleptonic widths.
As a result,\, there is a
sizeable difference ($\se 10\%$) between the above calculated and
the experimental numbers. A possible explanation may be due to the SU(3)
symmetry breaking effects which were neglected in these calculations.
At first sight,\, however,\, most of them tend to increase the discrepancy
rather than to decrease it. Clearly,\, this subject requires careful
investigation which is out the scope of this paper.

One of the important results of all the calculations above is that the
factorization approximation works well for the matrix elements,\, i.e. the
nonfactorizable contributions which have the same parametrical behaviour at
large $M_Q$ are,\, in comparison,\,  an order of magnitude smaller.
Nevertheles,\, they are of importance due to specific features of the problem
considered. Firstly,\, these nonfactorizable parts enter
the effective Lagrangian with much larger coefficients. Secondly,\, the
factorizable contributions are additionally suppressed by the smaller phase
space (see point 3 above).

Let us add a few words about baryon lifetimes. It seems clear
that the pattern here is very unlike those for the
pseudoscalar mesons. The entire structure of the matrix elements of $L_{eff}$
is quite different,\, and the reasons connected with the helicity
suppression of factorizable contributions are not operative here. So,\, one
can expect that: 1) the scale of the 4-fermion operator contributions to
the baryon lifetimes is potentially a few times larger (term by term)
than those for the pseudoscalar mesons;\, 2) the significance
of nonfactorizable contributions will be smaller for baryons. Therefore,\,
it seems,\, the main problem here will be to calculate reliably the
factorized parts of the matrix elements.

For the B mesons,\, all the above qualitative properties remain true but
,\, clearly,\, the role of all power corrections to the Born term becomes
much smaller. It seems clear now also that the
factorization approximation will be especially good for the $B_c$ mesons.

It is important that we understand now the properties of all the main
contributions giving rise the lifetime differences and that there is
sufficiently good agreement between the calculations for the D mesons and
the experimental data. This gives us confidence that the predictions
obtained above for the B mesons are reliable. Therefore,\, we can insist
now that the lifetime difference between the $B^{\pm}$ and $B^{o}$ mesons
will not exceed $\se 5\%$\,, while those between the $B^o$ and $B_s$
mesons have to be even smaller. As it was pointed out above,\,
the scale of the 4-fermion operator contributions for the b-baryons can be
a few times larger.

Surprisingly large corrections to the factorization approximation ($\se -18\%$)
are found for $B^o-{\bar B}^o$ mixing. This reduction of the $B_d-{\bar B}_d$
mixing mass is of importance,\, in particular,\, for the
parameters of the unitarity triangle. These latter are calculated in sect.15
and are:
\bq \lambda\se 0.22\,,\quad A\se 0.825\,,\quad \rho\se -0.40\,,\quad
\eta\se 0.20\,,\eq
\bq \sin (2\alpha)\se 0.60\,,\quad \sin (2\beta)\se 0.28\,,\quad
\sin(2\gamma)\se -0.80\,.\eq

We would like to emphasize finally that the experimental value of
$Br(D\ra e\nu+X)$
requires a noticeably larger value of the c-quark mass ($M_c\se 1.65\,GeV)$,\,
in comparison with those $(M_c\se 1.4-1.5\,GeV)$ in common use. As a
result,\, because the mass formulae tell us that the quark mass difference
is close to those of the mesons,\, this leads to a value of the b-quark mass
$(M_b\se 5.04\,GeV)$ which is also considerably larger than those in common
use $(M_b\se 4.6-4.8\,GeV)$. Further,\, the chain of the argument is as
follows. The large value of $M_b$ leads to a small value for $f_B\,:
f_B(M_B)\se 113\,MeV$. Together with (see sect.14) $B_B\se 0.82$,\, this
requires a large value of the t-quark $\overline {MS}$-mass: $M^*_t
\se 180\,GeV$,\, to reproduce the experimental value of $x_d\se 0.70$. This
results finally in the rather small value of $\epsilon^{\prime}/
\epsilon \sim \left (\pm 1\cdot 10^{-4}\right )$,\, which is incompatable
with the NA31-group result but does not contradict to those from the
E731-group.

\begin{center} {\bf Acknowlegements} \end{center}

I would like to thank J.\,Soffer for useful discussions on the subject of this
paper.

Parts of this work were made during my stays at the Centre de Physique
Theorique of Marseille University and the Physics Department of the
Weizmann Institute. I express my deep gratitute to both staffs and
especially to J.\,Soffer for the kind hospitality extended to me.

This work was supported in part by Grant $N^o$ RAK000 from the International
Science Foundation.

\newpage

\begin {thebibliography}{99}

\bi{FM} H.Fritzsch, P.Minkovsky, Phys.Lett. {\bf B90}(1980)455

\bi{BWS} W.Berneuther, O.Nachtman, B.Stech, Z.Phys. {\bf C4}(1980)257

\bi{BSS}  M.Bander, D.Silverman, A.Sony, Phys.Rev.Lett., {\bf 44}(1980)7

\bi {Gub}  B.Guberina, S.Nussinov, R.D.Peccei, R.Ruckl,
           Phys.Lett. {\bf B89}(1981)111

\bi {VS0} M.Voloshin, M.Shifman, 1982, in V.Khoze, M.Shifman, Uspekhi
          Fiz. Nauk {\bf 140}(1983)3

\bi {BGT} N.Bilik, B.Guberina, J.Trampetic, Nucl.Phys. {\bf B248}(1984)26

\bi {VS} M.Voloshin, M.Shifman, Yad. Fiz. {\bf 41}(1985)187;  ZhETF {\bf 91}
            (1986)1180

\bi {BU} I.I.Bigi, N.G.Uraltsev, Phys.Lett. {\bf B280}(1992)120

\bi {BUV1} I.I.Bigi, N.G.Uraltsev, A.I.Vainshtein, Phys.Lett.
           {\bf B280}(1992)430; {\bf B297}(1993)477(E)

\bi {BS1} B.Blok, M.A.Shifman, Nucl.Phys. {\bf B399}(1993)441

\bi {LS} M.Luke, M.Savage, Phys.Lett. {\bf B321}(1994)88

\bi {Big} I.I.Bigi, N.G.Uraltsev, Preprint CERN-TH-7063/93

\bi {LN} Z.Ligeti, Y.Nir, Preprint WIS-94/2/Jan-PH

\bi {SUV} M.Shifman, N.G.Uraltsev, A.I.Vainshtein, Preprint TPI-MINN-94
/13-T

\bi {IW} N.Isgur, M.Wise, in {\bf B decays}, S.Stone, Ed. (World
Scientific) 1992

\bi {Ge} H.Georgy, In Proceedings of 1991 Theoretical Advanced Study
Institute, eds. R.K.Ellis, C.T.Hill, J.D.Lykken, World Scientific,
Singapore 1992

\bi {Neu} M.Neubert, Preprint SLAC-PUB-6263, 1993

\bi {BSUV} I.I.Bigi, M.Shifman, N.G.Uraltsev, A.I.Vainshtein, Preprint
CERN-TH-71/94

\bi {BB} M.Beneke, V.M.Braun, Preprint MPI-PhT/94-9

\bi {BBZ} M.Beneke, V.M.Braun, V.I.Zakharov, Preprint MPI-PhT/94-18

\bi {A1} G.Altarelli et.al., Nucl.Phys. {\bf B208}(1982)365

\bi{VSG} M.Voloshin, M.Shifman, Yad.Fiz. {\bf 45}(1987)463

\bi {PW} H.Politzer, M.Wise, Phys.Lett. {\bf B206}(1988)681

\bi{Ioffe} V.M.Belyaev, B.L.Ioffe, ZhETF {\bf 83}(1982)876

\bi {PDG} Particle Data Group, Phys.Rev. {\bf D45}(1992)S1

\bi {MD} M.Danilov, Preprint ITEP 92-93, Moscow;\\
	 DELPHI, Preprint CERN-PPE/94-24;\\
        ALEPH, Preprint CERN-PPE/94-17

\bi {Beh} R.E.Behrends et al., Phys.Rev. {\bf 101}(1956)866

\bi {CM} N.Cabibbo, L.Maiani, Phys.Lett {\bf B79}(1978)109

\bi {PBB} P.Ball, V.Braun, Phys.Rev. {\bf D49}(1994)2472

\bi{CZ5} V.Chernyak,\, A.Zhitnitsky, Phys. Rep. {\bf 112}(1984)173

\bi {Sh} E.V.Shuryak, Nucl.Phys. {\bf B198}(1982)83

\bi {CZ1} V.Chernyak, A.Zhitnitsky, I.Zhitnitsky, Yad.Fiz.
          {\bf 38}(1983)1277

\bi {AE} T.M.Aliev, V.L.Eletsky, Yad.Fiz. {\bf 38}(1983)1537

\bi {CZ2} V.Chernyak, I.Zhitnitsky, Nucl.Phys. {\bf B345}(1990)137

\bi {Br} J.Broadhurst, Phys.Lett, {\bf B101}(1981)423

\bi {BG1} J.Broadhurst, S.C.Generalis, Phys.Lett. {\bf B139}(1984)85

\bi {BG2} J.Broadhurst, A.G.Grozin, Phys.Lett. {\bf B274}(1992)421

\bi {BS2} B.Blok, M.A.Shifman, Preprint TPI-MINN-93/55-T

\bi {Cor} J.Cortes, X.Pham, A.Tounsi,  Phys.Rev. {\bf D25}(1982)188

\bi {A2} J.Altarelli, S.Petrarca, Phys.Lett {\bf B261}(1991)303

\bi {PS} W.F.Palmer, B.Stech, Phys.Rev. {\bf D48}(1993)4174

\bi {BBSV} I.I.Bigi, B.Blok, M.Shifman, A.Vainshtein, Phys.Lett
                     {\bf B323}(1994)408
\bi {VUKS} M.Voloshin, N.Uraltsev, V.Khoze, M.Shifman,
           Yad.Fiz. {\bf 46}(1987)181

\bi {BJW}  A.J.Buras, M.Jamin, P.H.Weisz, Nucl.Phys. {\bf 347}(1990)491

\bi {BH} A.J.Buras, M.K.Harlander, Preprint MPI-PAE/PTh 1/92

\bi{CDF} The LEP Collaboration, Preprint CERN/PPE/93-157 \\
              F.Abe et al. (CDF), Preprint FermiLab-PUB-94/097-E

\bi{CB} C.Bernard, A.Sony, Proc. Lattice-89, eds.N.Cabibbo et al.,
                           Nucl.Phys. {\bf B17} (Proc. Suppl.) (1990)495\\
F.R.Brown et al., Columbia Collaboration, Lattice Workshope\\
                             in Tallahasse, 1990 \\
M.B.Gavela et al., Phys.Lett. {\bf B206}(1988)113;
                               Nucl.Phys. {\bf B306}(1988)677 \\
G.W.Kilcup et al., Phys.Rev.Lett. {\bf 64}(1990)25

\bi{BJL} A.J.Buras, M.Jamin, E.Lautenbacher, Preprint MPI-Ph/93-11

\bi{Flynn} J.M.Flynn, L.Randall, Phys.Lett. {\bf B224}(1989)221

\bi{Bur} H.Burckhardt et al. (NA31), Phys.Lett. {\bf B206}(1988)169

\bi{Pat} L.K.Gibbons et al. (E731), Phys.Rev.Lett. {\bf 70}(1993)1203

\bi{BW} G.Barr (NA31) and B.Winstein (E731), talks at the XV Int. Symp. on
Lepton Photon Inter., Geneva, July 1991

\bi{Vys}  A.I.Vainshtein et al., Yad.Fiz. {\bf 23}(1976)1024\\
	M.I.Vysotsky, Yad.Fiz. {\bf 31}(1980)1535

\bi{GW}  F.J.Gilman, M.B.Wise, Phys.Rev. {\bf D27}(1983)1128

\bi{C0} V.Chernyak, A.Zhitnitsky, Nucl.Phys. {\bf B201}(1982)492

\end{thebibliography}

\newpage

\begin{center} {\bf Appendix} \end{center}

For the convenience of the reader,\, we list below the notations and numerical
values of various parameters used in the text.\\

Parameters:\\
\bq M_b=5.04\,GeV\,[Eq.(103)]\,,\quad M_c=1.65\,GeV\,[sect.4]\,,
\quad \mu_o^2=0.5\,GeV^2\,[Eq.(43)];\eq
\bq \alpha_s(M^2_b)=0.204\,,\quad \alpha_s(M^2_c)=0.310\,,\quad \alpha_s(
\mu^2_o)=0.580\quad [Eq.(49)];\eq
\bq <0|\,{\bar q}i\gamma_5\,Q|D>=r_D=f_D\frac{M_D^2}{M_c}\,;\eq
\bq f_D(\mu_o)=144\,MeV\,,\quad f_D(M_c)=165\,MeV \quad [Eq.(123)];\eq
\bq  f_B(\mu_o)=89\,MeV\,,\quad f_B(M_b)=113\,MeV\quad [Eq.(128)];\eq
\bq \eta_{co}=\frac{\alpha_s(\mu_o^2)}{\alpha_s(M_c^2)}= 1.87\,,\quad
\tau_{co}=\eta_{co}^{1/2}=1.37\,;\eq
\bq \eta_{bo}=\frac{\alpha_s(\mu_o^2)}{\alpha_s(M_b^2)}=2.84\,,\quad
\eta_{bc}=\frac{\alpha_s(M_c^2)}{\alpha_s(M_B^2)}=1.52\,,\quad \tau_{bo}=
\eta_{bc}^{27/50}\,\tau_{co}=1.715\,;\eq
\bq \left (\,1-\eta_{co}^{-2/9}\,\right )=0.13\,,\quad  \rho_o=\frac{
\alpha_s(\mu_o^2)}{2\,\pi} \log {\frac{\mu_{max}^2}{\mu_{min}^2}}=0.1\,;\eq
\bq <{\vec p}^2_c>=\frac{<D|\,{\bar c}\,(i\,{\vec D})^2\,c\,|D>}{<D|\,
{\bar c}\,c\,|D>}=0.3\,GeV^2\,,\quad \Delta_K=\frac{<{\vec p}^2_c>}{2\,
M_c^2}\,;\eq
\bq \Delta_G=\frac{1}{M_c^2}\,\frac{<D|\,{\bar
c}\frac{i}{2}\,g_s\,\sigma_{\mu\,\nu}\,G_{\mu\,\nu}^a\,\frac{\lambda^a}
{2}\,c\,|D>}{<D|\,{\bar c}\,c\,|D>}=\frac{3}{4}\,\frac{M_{D^{\*}}^2-M_D^2}
{M_c^2}=0.15\,;\eq
\bq N_v=2.54\quad [Eqs.(51),(52)],\quad A=0.40\quad [Eqs.(73),(74)]\,.\eq

The matrix elements (sects.9,10):\\

\bq  <D(p)|\,{\bar c}\,\Gamma_\mu\,q\cdot {\bar q}\,\Gamma_\nu\,c\,|D(p)>_
{\mu_o}= p_\mu\,p_\nu\,f_D^2(\mu_o)\,\delta_V+{{2}\over{27}}\rho_o\,
g_{\mu\,\nu}\,r_D^2(\mu_o)\,, \eq
\bq <D(p)|\,{\bar c}\,\Gamma_\mu{{\lambda^a}\over{2}}\,q\cdot {\bar q}
{{\lambda^a}\over{2}}\Gamma_\nu\,c\,|D(p)>_{\mu_o}= \left
({{p_\mu\,p_\nu}\over{p^2}}-g_{\mu\,\nu}\right ) C_G\,
\delta_V+g_{\mu\,\nu}P_V\,\delta_V- & \nonumber \\
& \displaystyle \frac{1}{81}\,\rho_o\,g_{\mu\,\nu}r_D^2(\mu_o)\,, \eq
Here: $\delta_V\,$ is unity if the quark field in the matrix element has a
valence flavour and is zero otherwise.
\bq C_G^c=0.48\cdot 10^{-2}\,GeV^4\quad [Eq.(162)],
\quad C_G^b=1.15\cdot 10^{-2}\,GeV^4\quad [Eq.(163)]\,; \eq
\bq P_V^c=0.685\cdot 10^{-3}\,GeV^4\,,
\quad P_V^b=0.18\cdot 10^{-2}\,GeV^4\quad [Eq.(192)];\eq
\bq  <D(p)|\,(J_{\rho}^a\cdot {\bar c}\gamma_
\rho(1\pm \gamma_5){{\lambda^a}\over{2}}\,c\,|D(p)>_{\mu_o}=
-{{4}\over{9}}\,r_D^2(\mu_o\,)\left (1-{{M_c^2}\over{2\,
M_D^2}}\right )+\cdots \,,\eq
where the dots mean that we neglected corrections because this matrix element
appears itself as a correction in the above calculations.

In addition to the 4-quark operators considered in the text,\, there is also
the nonvalence 4-quark operator contribution originating from the Born
diagram,\, and we show below that it is negligibly small. Using the
Fock-Schwinger gauge:
\bq x_{\mu}A_{\mu}(x)=0, \quad A_{\mu}(x)= \frac{1}{2}x_{\nu}G_{\nu\,\mu}(0)+
\frac{1}{3}x_{\nu}x_{\lambda}D_{\lambda}G_{\nu\,\mu}(0)+\cdots\,, \eq
the Born diagram contribution can be represented in the form:
\bq  \Gamma_{Born} \sim <D(p)|\Bigl \{ {\bar c}(x){\stackrel {\leftarrow}
\partial_{\mu}}^2\,(1-\gamma_5)\cdot
\partial_{\nu}^2\,(\,i{\hat \partial})\,c(x)\Bigr \}_{x=0}|D(p)>\,.\eq
Using the equations of motion: $i{\hat \partial}c=M_c\,c-g_s{\hat A}\,c,$ it
is not difficult to obtain:
\bq \Gamma_{Born}\sim M_c^5<D(p)|{\bar c}(0)(1-\gamma_5)\left (1-ig_s\sigma_
{\mu\,\nu}G_{\mu\,\nu}(0)\right )\,c(0)|D(p)>- \nonumber & \\ &
\displaystyle \frac{1}{6}g_s^2\,M_c^2<D(p)|{\bar c}(0)\gamma_\mu (1+\gamma_5)
\frac{\lambda^a}{2}c(0)\cdot J^a_{\mu}(0)|D(p)>+\frac{1}{2}\delta_V+
\frac{1}{2}\delta_A+\cdots\,,\eq
where $"\cdots"$ denotes higher dimension terms. Also ($\stackrel
{\leftrightarrow} \partial_{\mu}=\stackrel {\ra} \partial_{\mu}-\stackrel
{\la} \partial_{\mu}$),
\bq \delta_V=<D(p)|{\bar c}(0)\gamma_\mu \frac{\lambda^a}{2}{\stackrel
{\leftrightarrow} \partial_\nu}c(0)\cdot G_{\mu\,\nu}^a(0)|D(p)>=0 \eq
due to C-parity,\, and
\bq \delta_A=<D(p)|{\bar c}(0)\gamma_\mu \gamma_5 \frac{\lambda^a}{2}
{\stackrel {\leftrightarrow} \partial_\nu}c(0)\cdot G_{\mu\,\nu}^a(0)
|D(p)>=0 \eq
due to P-parity. Therefore,\, the correction due to the 4-fermion operator is:
\bq \delta_4=1-\frac{2}{3}\pi\,\alpha_s(M_c)\frac{1}{2\,M_D\,M_c^3}<D|{\bar c
}(0)\gamma_\mu(1+\gamma_5)\frac{\lambda^a}{2}c(0)\cdot
J^a_{\mu}(0)|D>\,.\eq
This becomes after the factorization of the matrix element:
\bq \delta_4\se 1+\frac{4}{27}\pi\,\alpha_s(M_c)\frac{f_D^2(M_c)\,M_D^3}
{M_c^5}\left (1-\frac{M_c^2}{2\,M_D^2}\right )\se 1+1.3\cdot 10^{-3}\,,\eq
so that it is negligble even for the D-meson.

For the semileptonic decays the above described 4-quark operator contribution
is the only one,\, in addition to the ${\bar c}c,\, {\bar c}\sigma_{\mu\,\nu}
G_{\mu\,\nu}\,c$ and 4-quark operators described in the text,\, which is
capable to give $\sim O(\Lambda_{QCD}^3/M_c^3)$ correction. For the
nonleptonic decays,\, in addition,\, it is not difficult to see that it
is sufficient to replace $\Delta_G$ in the last term in Eq.(25) (which is due
to the fig.2 diagram) by its original expression:
\bq \Delta_G \ra -\, \frac{<D|{\bar c}\frac{\lambda^a}{2}\gamma_{\mu}\gamma
_5\,i{\stackrel {\leftrightarrow} \partial_{\nu}}\,c\cdot {\widetilde G_{\mu\,
\nu}}^a|D>}{2\,<D|{\bar c}c|D>\,M_c^3}\,.\eq

\newpage

\begin{center} {\bf Figure captions} \end{center}

Fig.1\quad The Born contribution \\

Fig.2\quad The soft gluon emission giving rise the correction $0(1/M_Q^2)$\\

Fig.3\quad The diagrams contributing into the 4-fermion operators\\

Fig.4\quad The nonvalence factorizable penguin contribution (PNV)\\

Fig.5\quad The diagrams contributing to the sum rule Eq.(118) \\

Fig.6a\quad The nonfactorizable gluon contribution to the weak annihilation\\

Fig.6b\quad The same for the cross weak annihilation    \\

Fig.7\quad  The diagrams for the meson transition into a current\\
  \hspace*{3cm}   with an emission of a nonperturbative gluon \\

Fig.8\quad  The nonfactorizable contribution to the matrix element
     $<D|\,S_{\mu\,\nu}\,|D>$ \\

Fig.9\quad  The nonvalence nonfactorizable penguin contribution (NV)\\

Fig.10\quad The valence nonfactorizable penguin contribution  \\

\end{document}